%% file: main.tex
\documentclass[acmlarge]{acmart}

\AtBeginDocument{%
  \providecommand\BibTeX{{%
    \normalfont B\kern-0.5em{\scshape i\kern-0.25em b}\kern-0.8em\TeX}}}

\usepackage{listings}

\usepackage{url}

\usepackage{breakurl}

\definecolor{commentgreen}{rgb}{0.3,0.5,0.5}
\definecolor{keyred}{rgb}{0.63,0.129,0.258}
\definecolor{codegray}{rgb}{0.5,0.5,0.5}
\definecolor{codepurple}{rgb}{0.58,0.4,0.82}
\definecolor{backcolour}{rgb}{0.95,0.95,0.92}
\definecolor{maroon}{cmyk}{0,0.87,0.68,0.32}

\lstdefinestyle{mystyle}{
    commentstyle=\itshape\color{commentgreen},
    keywordstyle=\bfseries\color{keyred},
    numberstyle=\tiny,
    stringstyle=\color{codepurple},
    basicstyle=\ttfamily\footnotesize,
    breakatwhitespace=false,         
    breaklines=true,                 
    captionpos=b,                    
    keepspaces=true,                 
    numbers=left,                    
    numbersep=5pt,                  
    showspaces=false,                
    showstringspaces=false,
    showtabs=false,                  
    tabsize=2,
    frame={bottomline}
}
\begin{document}

\title{A Survey on EOSIO Systems Security: Vulnerability, Attack, and Mitigation}

\author{Ningyu He}
\affiliation{ 
      \institution{Peking University}
      \city{Beijing} 
      \country{China}
}
\email{ningyu.he@pku.edu.cn}

\author{Haoyu Wang}
\affiliation{ 
      \institution{Huazhong University of Science and Technology}
      \city{Hubei} 
      \country{China}
}
\email{haoyuwang@hust.edu.cn}

\author{Lei Wu}
\affiliation{ 
      \institution{Zhejiang University}
      \city{Zhejiang} 
      \country{China}
}
\email{lei_wu@zju.edu.cn}

\author{Xiapu Luo}
\affiliation{ 
      \institution{The Hong Kong Polytechnic University}
      \city{Hong Kong} 
      \country{China}
}
\email{csxluo@comp.polyu.edu.hk}

\author{Yao Guo}
\affiliation{ 
      \institution{Peking University}
      \city{Beijing} 
      \country{China}
}
\email{yaoguo@pku.edu.cn}

\author{Xiangqun Chen}
\affiliation{ 
      \institution{Peking University}
      \city{Beijing} 
      \country{China}
}
\email{cherry@sei.pku.edu.cn}

\renewcommand{\shortauthors}{Ningyu, et al.}

\begin{abstract}
EOSIO, as one of the most representative blockchain 3.0 platforms, involves lots of new features, e.g., delegated proof of stake consensus algorithm and updatable smart contracts, enabling a much higher transaction per second and the prosperous decentralized applications (DApps) ecosystem. According to the statistics, it has reached nearly 18 billion USD, taking the third place of the whole cryptocurrency market, following Bitcoin and Ethereum. Loopholes, however, are hiding in the shadows. EOSBet, a famous gambling DApp, was attacked twice within a month and lost more than 1 million USD. No existing work has surveyed the EOSIO from a security researcher’s perspective. To fill this gap, in this paper, we collected all occurred attack events against EOSIO, and systematically studied their root causes, i.e., vulnerabilities lurked  in all relying components for EOSIO, as well as the corresponding attacks and mitigations. We also summarized some best practices for DApp developers, EOSIO official team, and security researchers for future directions.
\end{abstract}



\maketitle

\input{Section-Intro.tex}

\input{Section-Background.tex}

\input{Section-Vulnerability.tex}

\input{Section-Attack.tex}

\input{Section-Mitigation.tex}

\input{Section-Discussion.tex}

\input{Section-Related.tex}

\section{Conclusion}
In this paper, we detail all the vulnerabilities related to EOSIO that have appeared in history. These vulnerabilities cover the three levels of EOSIO, which are smart contracts, EOS VM, and blockchain itself. 
In addition to this, we demonstrate the attack techniques for most of the vulnerabilities, as well as the mitigations (and the best practices in programming) for those that cannot be officially fixed.
For DApp developers, EOSIO official team and security researchers, we summarize some recommendations to indicate the direction ahead. 
As far as we know, this is the first survey against EOSIO. Although there have been many surveys against Ethereum's vulnerabilities and attacks, the mechanism of EOSIO and Ethereum is very different.
Therefore, our work can be of great help to EOSIO smart contract developers and researchers.

\bibliographystyle{ACM-Reference-Format}
\bibliography{citation.bib}


\end{document}

%% file: Section-Intro.tex
\section{Introduction}
\label{sec:intro}

When Satoshi Nakamoto released the whitepaper of Bitcoin in 2008~\cite{bitcoin}, cryptocurrency has become one of the most popular topics in the area of computer science and finance. By the time of writing, Bitcoin is still the leading one, far beyond the second place, of numerous kind of cryptocurrencies, with the market capitalization of more than 134 Billion USD~\cite{bitcoin-cap}.
As the Bitcoin's follower, Ethereum, who takes the second place, brings in the concept of \textit{smart contract}, in which the programmer are allowed to interact with the data stored in blockchain by the means of programmable script.

However, these early-staged implementation has a fatal limitation: a poor scalability. The Bitcoin can only theoretically support 27 transactions per second (TPS) at most~\cite{bitcoin-tps}; while Ethereum can achieve approximately 50 TPS~\cite{ethereum-tps}.
Moreover, the Ethereum is still in ongoing development, the frequently adding and deleting of features in its programming language lead to developers' unfamiliarity and unexpected behaviors.
Lots of works and technical blogs~\cite{atzei2017survey,peckshield-blog,slowmist-zone} have indicated kinds of vulnerabilities existed in Ethereum smart contract and its virtual machine. At the meanwhile, several works~\cite{wang2020contractward,gao2019easyflow,grech2018madmax,torres2018osiris} focused on the vulnerabilities' detection in a variety of ways.
Therefore, lots of competitors emerged, one of the most influential products is EOSIO~\cite{eos}.

EOSIO launched its initial coin offering~\cite{ico} (ICO) in 2017, and raised more than 4 billion USD which is the largest one ever~\cite{eos-ico}. Its main purpose is to be Ethereum alternative, thus EOSIO improves upon intrinsic shortcomings of Ethereum, e.g., scalability and speed.
According to a EOSIO network monitor~\cite{eos-monitor}, it have reached around 4,000 TPS in history, which is tenfolds higher than Ethereum's and Bitcoin's.
Except for its high TPS, EOSIO choose C++ as its smart contract language, and WebAssembly~\cite{webassembly} (Wasm) as its target language executed in EOS Virtual Machine (EOS VM). Though Wasm is relatively younger than C++, it is endorsed by several influential technical companies. Additionally, Wasm aims to execute at native speed by taking advantage of common hardware capabilities.
Therefore, the high TPS and efficiency of EOSIO smart contract bring EOSIO a huge success. For example, EOSIO has successfully surpassed Ethereum in decentralized applications (DApp) transactions just three months after its launch in June 2018~\cite{eos-surpass-eth}, and further increased dozens of times after another several months.

However, due to developer's unfamiliarity of developing EOSIO smart contract, kinds of vulnerabilities has been introduced. For example, EOSBet~\cite{eosbet}, one of the most famous gambling DApp in EOSIO, have been attacked twice in just one month~\cite{eosbet-attack-fake-eos,eosbet-attack-fake-receipt} due to the lack of verification of two different key arguments in transaction respectively. In these two attacks, EOSBet has lost nearly 190 thousand of EOS, which was more than 1 million USD according to the exchange rate then.
According to several blockchain security companies' blogs, forums and official announcements, we have counted 113 attacks against several kinds of vulnerabilities just within one year after EOSIO launched. 

Except for vulnerabilities imported in smart contract by developers, EOS VM, the stack-based machine used to execute smart contract, also has some inherent flaws that are used by some attackers~\cite{eosvm-asset-overflow, inline-reflex}.
Additionally, EOSIO utilizes an energy-saved and efficient consensus algorithm, delegated proof-of-stake~\cite{dpos} (DPoS), as its consensus algorithm instead of widely-used proof-of-work (PoW). Lots of novel characteristics are introduced by DPoS and EOSIO new transaction model, e.g., block producer, action and transaction. Some of them can result in unexpected behaviors which are not considered by the official.

This paper makes the following contributions:
\begin{enumerate}
	\item We present the first systematic exposition of the vulnerabilities in EOSIO ecosystem, ranging from smart contracts to the consensus algorithm. The hierarchical presentation of vulnerabilities clearly illustrates the direct causes of these loopholes which can be referenced by DApp developers and security researchers; 
	\item We comprehensively survey all existing attack events against all vulnerabilities, as well as their direct consequence, attack tactics and corresponding countermeasures;
	\item We well study all existing mitigations on identifying vulnerabilities or happened attack events, including program analysis and transaction analysis. Moreover, we propose a set of best practices for developers to avoid financial losses as much as possible.
\end{enumerate}

The whole paper is organized as follows: \S\ref{sec:background} depicts the whole picture of EOSIO blockchain platform and performs a comparison between Ethereum smart contract and EOSIO's on bytecode level. Then, \S\ref{sec:vul} and \S\ref{sec:attack} introduce the underlying principles of vulnerabilities and attacks, which are collected from several well-known sources, like~\cite{peckshield-blog, slowmist-zone, eos-tech-blog}, respectively. Moreover, \S\ref{sec:mitigation} summarizes best practices and current existing work whose targets are all against security issues of EOSIO smart contracts. Finally, we propose some best practices in \S\ref{sec:discussion} to DApp developers, EOSIO official team and security researchers according to the observation from the previous three sections, and \S\ref{sec:related} summarizes existing related work from several perspectives.

%% file: Section-Background.tex
\section{Background}
\label{sec:background}
\begin{figure}[tbp]
\centerline{
\!\!\!\!\!\!\!\!\!\!\!\!\!\!\!\!\!\!\!\!\!\!\!\!\!\!
\includegraphics[width=0.8\textwidth]{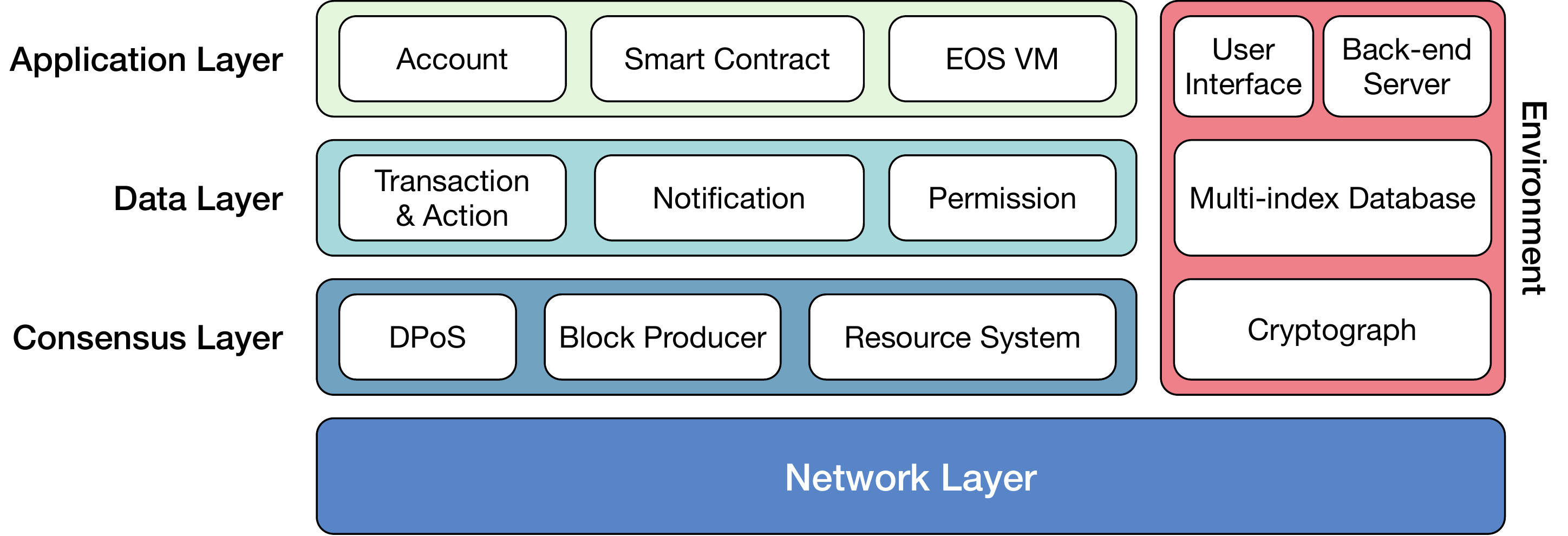}
}
\caption{The architecture of EOSIO.}
\label{fig:structure-eosio}
\end{figure}

In this section, we will briefly introduce the key concepts related to the EOSIO platform to facilitate the understanding of this work. 
As shown in Fig~\ref{fig:structure-eosio}, according to their dependency, we divide the components of EOSIO into five categories: \textbf{\textit{application layer}}, \textbf{\textit{data layer}}, \textbf{\textit{consensus layer}}, \textbf{\textit{network layer}} and \textbf{\textit{environment}}.
In application layer, EOS Virtual Machine (EOS VM) runs on each nodes in EOSIO, responsible for executing smart contracts that are invoked by other accounts.
All these interactions between accounts are achieved by transactions (and actions) and notifications with their carrying permission, which are all packaged and stored at data layer.
Moreover, as we mentioned in \S\ref{sec:intro}, EOSIO significantly outperforms Bitcoin and Ethereum in terms of Transactions Per Second (TPS) and prospers in DApp ecosystem. This is mainly due to its adopted consensus algorithm and resource system, which will be introduced in consensus layer.
As an auxiliary component, environment offers several key technology to serve the above three layers, e.g., a user-friendly front-end interface and a trustworthy back-end serve can support the application layer.
Last, blocks, as the basic unit that will be propagated across all nodes, are created and relayed in the network layer, played as infrastructure for all these stuff.
For full details, readers can refer to the official documents (e.g., \cite{eos-whitepaper}) for a systematic illustration.

\subsection{Application Layer}
\label{sec:background:application}

\subsubsection{Account}
\label{sec:bac:app:account}
Instead of the unreadable and random-like address adopted by Ethereum and Bitcoin to identify accounts, the account name of EOSIO is composed of at most 12 characters. And EOSIO requires the uniqueness of all accounts' name to avoid identity impersonation.
Moreover, unlike the distinction between External Owned Account (EOA) and smart contract in Ethereum~\cite{eoa-vs-sc}, there is only one type of account in EOSIO, which can be regarded as a normal account and a smart contract simultaneously.

\subsubsection{Smart Contract}
\label{sec:bac:app:sc}
As we mentioned in \S\ref{sec:bac:app:account}, an account can be seen as a smart contract sometimes. To be specific, if a piece of compiled smart contract bytecode is stored in one's account, the account performs pre-defined logic written in the contract once it is invoked. 
Similar to one's balance, EOSIO smart contract is \textit{updatable}.
The mutability of smart contract not only allows the developers to patch loopholes, but also introduces users' concern that the unexpected behaviors may happen if they are not aware of the update.
EOSIO smart contracts are written in C++, and compiled to WebAssembly (Wasm)~\cite{webassembly}, a well-structured assembly-like language.

\subsubsection{EOS VM}
As with Ethereum, EOSIO adopts a virtual machine to execute the smart contracts, named as EOSIO Virtual Machine (EOS VM), embedded in each nodes of EOSIO.
EOS VM is a stack-based machine, i.e., all the operands and operators are pushed into the popped from a \texttt{stack}. Moreover, EOSIO supports another two sections: \texttt{local} and \texttt{global}, to share data within a function or across functions, respectively.
Furthermore, EOS VM provides a random-accessible linear array called \texttt{memory} to store the unbounded and non-permanent data, e.g., a string that will be printed out.

\subsection{Data Layer}
\label{sec:background:data}
Block is the basic unit that will be appended to EOSIO blockchain, the decentralized ledger.
A block consists of two key components: block header and block body. The former one contains some meta data, e.g., block height, timestamp; the latter one is composed of \textit{transactions}.
In EOSIO, a transaction can be further divided into \textit{actions}, in which it carried the invoker's permission to authorize behaviors. Moreover, EOSIO introduces a \textit{notification} mechanism that enables message delivering between accounts.

\subsubsection{Transaction \& Action}
\label{sec:bac:data:tx}
Each transaction in EOSIO is packed as a \textit{transaction instance}.
Moreover, each transaction is composed of two components: \textit{transaction header} and a list of \textit{action instances}. To be specific, the former one contains meta data, like transaction ID and its belonged block ID prefix; the latter one is a string of linked action instances.
Each action instance corresponds to an atomic behavior in EOSIO, consisting of four components: \textit{account name}, \textit{action name}, \textit{action data}, and \textit{authorization list}. The first three respectively represent the invoked account name, function name and concrete data to invoke the function; while the last one is the invoker's authorization to the current action instance (detailed in \S\ref{sec:background:data:permission}).

\begin{figure}[tbp]
\centerline{\includegraphics[width=0.8\columnwidth]{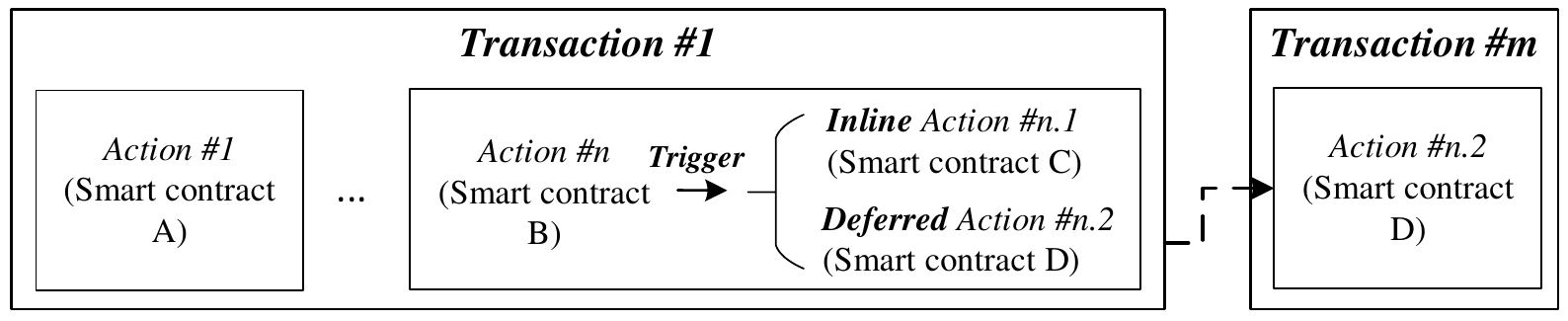}}
\caption{The model of transaction and action in EOSIO.}
\label{fig:tx-action}
\end{figure}

Moreover, actions can be created explicitly to invoke a smart contract, or generated implicitly as the side effect alongside other actions. Thus, actions are divided into two categories: \textit{explicit action} and \textit{implicit (inline) action}.
Explicit action is represented as an outermost and isolated action instance, like the \textit{action \#1} and the \textit{action \#n} in Fig.~\ref{fig:tx-action}; and implicit action shares the same context to which its parent action belongs. For example, the inline action \textit{action \#n.1} is the side effect when executing \textit{action \#n}, thus these two actions share an action instance.
Note that both explicit and implicit actions are executed in synchronous mechanism, it is necessary, however, for EOSIO to support the asynchronous execution because a single transaction has to be finished within 30 ms~\cite{eos-tx-30ms} to avoid the congestion.
Therefore, EOSIO allows an action to invoke a \textit{deferred action} as the \textit{action \#n.2} in Fig.~\ref{fig:tx-action}. The deferred action will be embedded in a future transaction after the delay period set by its initiator.

Similar to the Ethereum's rollback mechanism, any failure within a transaction will lead to its reversion, but it will not affect other already on-chained transactions.
For example, if \textit{action \#n.1} failed, the whole \textit{transaction \#1} will be reverted; if \textit{action \#n.2} in \textit{transaction \#m} fails, the \textit{transaction \#m} will be reverted but the \textit{transaction \#1} will keep everything intact.

\begin{figure}[tbp]
\centerline{\includegraphics[width=0.8\columnwidth]{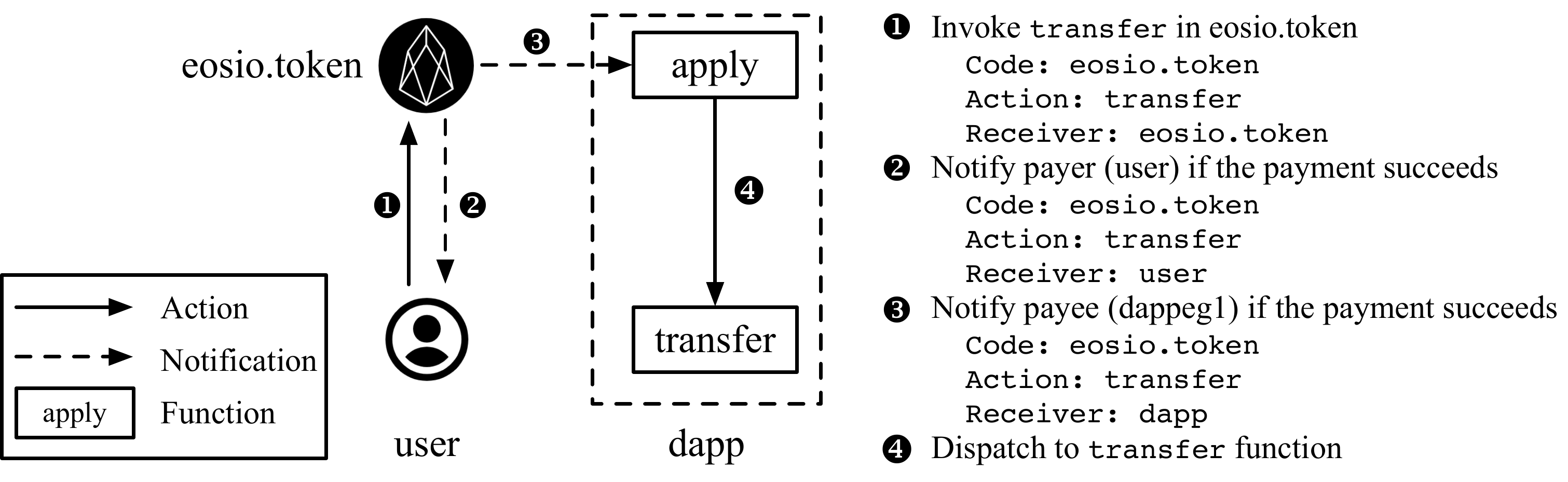}}
\caption{The process of transferring EOS from \texttt{user} to \texttt{dapp}.}
\label{fig:transfer-eos}
\end{figure}

\subsubsection{Notification}
\label{sec:background:data:notification}
EOSIO imports the \textit{notification} mechanism to enable message delivering between accounts.
Specifically, a smart contract can initiate a notification by \texttt{require\_recipient( )} within a function.
For example, the native token, dubbed \textit{EOS}, is issued by an official account: \texttt{eosio.token}, in which it maintains a balance table for all holders.
Therefore, as depicted in Fig.~\ref{fig:transfer-eos}, if the \texttt{user} intends to transfer EOS to the \texttt{dapp}, he has to request the \texttt{transfer} function in \texttt{eosio.token} firstly to update the corresponding rows of balance table. Once the update finished, \texttt{eosio.token} will \textit{notify} both of the payer and payee that the following logic can be performed continously (step 2 and 3).
Received notification is handled by a fixed entry-point function: \texttt{apply}, which must be implemented by each smart contract and will be detailed in \S\ref{sec:bac:data:apply}.

\begin{figure}[tbp]
\centerline{\includegraphics[width=0.6\columnwidth]{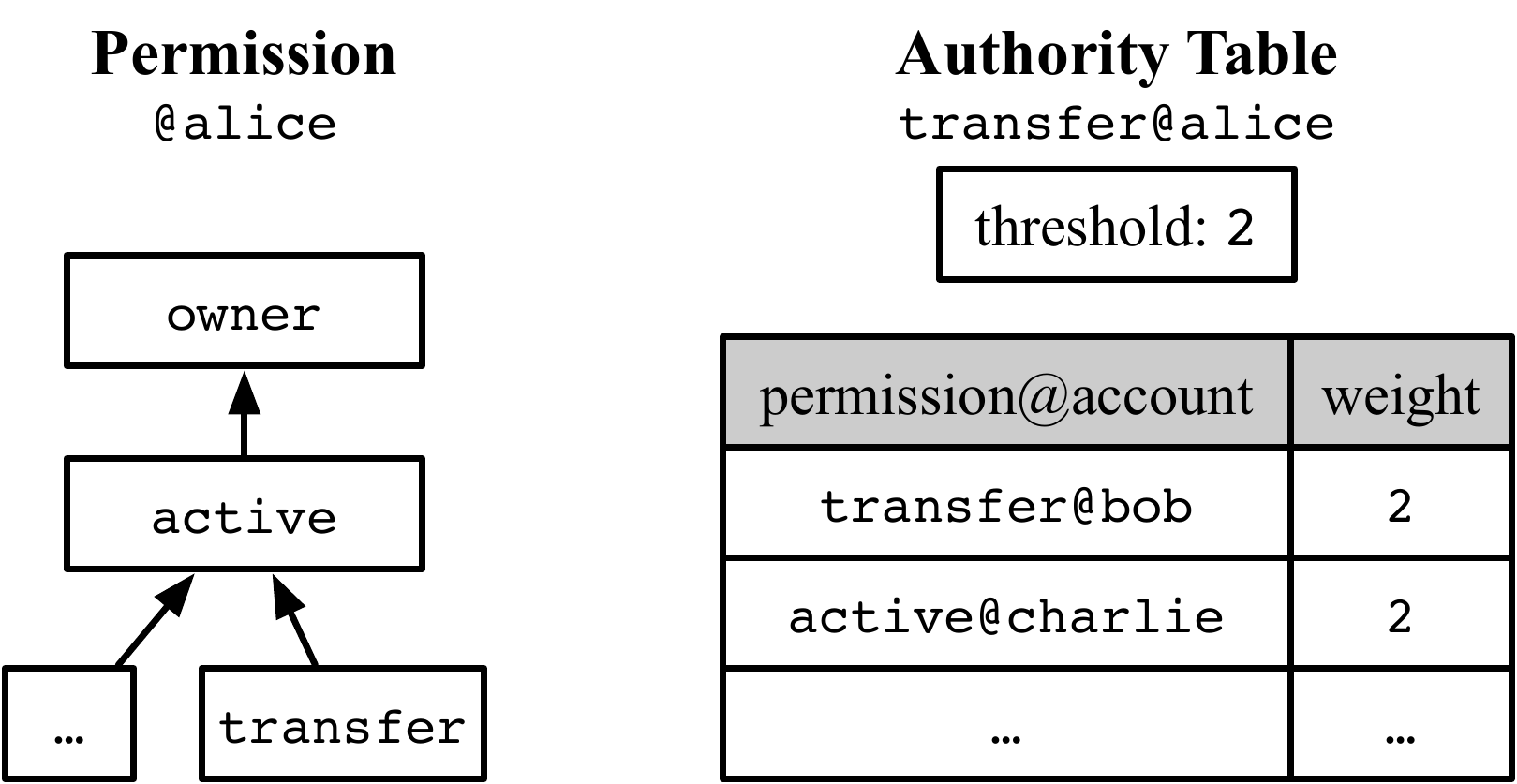}}
\caption{The hierarchical permissions of alice, and the corresponding authority table of permission transfer.}
\label{fig:permission-authority}
\end{figure}

\subsubsection{Permission System}
\label{sec:background:data:permission}
In EOSIO, each account is associated with a permission system. An concrete example, the permission system of \texttt{alice}, is shown in Fig.~\ref{fig:permission-authority}
Specifically, the permission system follows hierarchical structure, and there are always two default permissions: \texttt{owner} and \texttt{active}. Specifically, \texttt{owner} sits at the root of the permission hierarchy and correspond to one's highest permission; and \texttt{active} permission can do anything except changing the key associated with current account (controlled by the \texttt{owner}).
Following the hierarchical structure, accounts are allowed to create new permissions, like the \texttt{transfer} under the \texttt{active} permission.
Moreover, each permission is linked to an \textit{authority table}, in which it specifies the threshold that should be satisfied at least to perform the action which requires the permission, and the others' permission with their weights.
Within each action instance (see \S\ref{sec:bac:data:tx}), the initiator has to designate the carrying permission and signed the action by the corresponding key of the permission.
For example, if a function requires \texttt{transfer@alice}, the action with either \texttt{transfer@bob} or \texttt{active@charlie} could satisfy the requirement, as both of them are authorized by \texttt{transfer@alice}.

\begin{table}[t]
\centering
\caption{The meaning of arguments in \texttt{apply} under two circumstances.}
\begin{tabular}{ccc}
\toprule	
\textbf{}         & \textbf{Direct Invocation}  & \textbf{Notification}                                                                     \\ \midrule
\texttt{receiver} & The callee                  & The notified account                                                                      \\
\texttt{code}     & Current account             & The account initiating the notification         \\
\texttt{action}   & The invoked function's name & In which function the notification is initiated \\ \bottomrule
\end{tabular}
\label{table:arguments-apply}
\end{table}

\subsubsection{Dispatcher -- An Example}
\label{sec:bac:data:apply}
In EOSIO, there is a \textit{dispatcher}, named \texttt{apply}, responsible for handling the received action invocations and notifications. Moreover, its name and signature is fixed as shown in Listing~\ref{lst:apply-example}.
\texttt{apply} takes three arguments as input: \texttt{\textit{receiver}}, \texttt{\textit{code}}, and \texttt{\textit{action}}.
The \texttt{receiver} is the recipient; the \texttt{action} is the function's name; and the \texttt{code} indicates in which the code is actually executed. The concrete meanings of these three arguments under two circumstances: direct invocation or notification are shown in Table~\ref{table:arguments-apply}.
Therefore, if Listing~\ref{lst:apply-example} is the \texttt{apply} instance of \texttt{dapp} in Fig.~\ref{fig:transfer-eos}, it only accepts the notification from official account \texttt{eosio.token} (where \texttt{code} is \texttt{eosio.token}\footnote{The \texttt{N()} in L2 is the native string encoding function implemented by EOSIO, which will not be explained in the following.}) or a direct invocation (where \texttt{code} is itself). 

\begin{lstlisting}[language={C++}, caption={An example of \texttt{apply} in EOSIO smart contract. Note that this is vulnerable to the fake EOS vulnerability.}, label={lst:apply-example}, mathescape=true]
void apply(uint64_t receiver, uint64_t code, uint64_t action) {
  if(code == _self || code == N(eosio.token)) {
    switch(action) {
      // dispatches to corresponding funciton
    }
  }
}
\end{lstlisting}

\subsection{Consensus Layer}
\label{sec:background:consensus}

\subsubsection{DPoS \& Block Producer}
\label{sec:background:consensus:dpos}
The correctness of blockchain is essential for user's financial security, thus how to guarantee the data consistency across such a distributed system is the overriding issue for cryptocurrency.
Unlike Bitcoin and Ethereum, who utilizes traditional energy-consumed PoW consensus algorithm, EOSIO adopts \textit{delegated proof of stake (DPoS)}, a more efficient and environment-friendly one.
Specifically, all the accounts in EOSIO network are entitled to delegate their rights to someone whom they entirely believe in, and only the top 21 trusted accounts are called \textit{block producer} (BP). These 21 BPs are responsible for constructing blocks in turn.
Once a BP is on duty, it is responsible for verifying the correctness of the signature and executing actions in the transaction. After all the actions are executed without any failure, the transaction will be appended to a new block's body, which will then be broadcasted and validated by other nodes. While the block is validated by a supermajority of BPs ($> 2/3$), the block, including the transactions inside, will be irreversible.
Note that the order of producing blocks is negotiated in advance among those BPs, thus competitive mining process is unnecessary and deprecated in DPoS.

\subsubsection{Resource Model}
\label{sec:background:consensus:resource}
Similarly to other blockchain platforms, executing transactions and constructing blocks charges fee from initiators to prevent denial of service attack.
However, EOSIO does not calculate fee by length of data (e.g., Bitcoin)~\cite{btc-fee} or executed opcode (e.g., Ethereum)~\cite{eth-fee}, it introduces three key concepts to represent resources: \textit{NET}, \textit{CPU} and \textit{RAM}.
Specifically, all these three types of resources are provided by BPs, i.e., transaction validators and block constructors. NET is used to measure the amount of data that can be sent within a transaction, and CPU limits the maximum execution time of a transaction. These two are jointly called \textit{bandwidth}.
RAM limits the maximum space that can be occupied to store permanent data\footnote{The data is stored in the BPs' RAM to accelerate the retrieving process.}.
An account can exchange NET and CPU by mortgaging EOS, but has to buy RAM due to its scarcity. Users are free to use the bandwidth resources as long as there are available ones, i.e., not be mortgaged by any other accounts. However, the price of mortgaging bandwidth resources fluctuates, depending on how much EOS are staked in total.

\subsection{Network Layer}
\label{sec:background:network}
EOSIO also adopts P2P network to broadcast messages among the nodes distributed across the world.
In this work, however, we mainly focus on the vulnerabilities existed in its above layers: application, data and consensus layers. Therefore, the network layer is a black box for us in this work.
For more details about EOSIO network layer, e.g., \textit{block and transaction propagation} and \textit{P2P protocol}, please refer to the official documentation~\cite{eos-whitepaper}.

\subsection{Environment}
\label{sec:background:environment}
For all these layers mentioned above, it is necessary for them to be supported by some infrastructures.
To be specific, for DApp developers on the application layer, they need to provide user-friendly \textit{front-end interfaces} and \textit{back-end servers} to perform the corresponding functionalities, e.g., random number generation.
For items stored in the data layer, they have to be indexed in a \textit{multi-indexed database} to conduct an efficient CRUD\footnote{Create, Read, Update, and Delete} operations.
Last, the underlying \textit{cryptography} plays an important role for the consensus layer, like validating transactions and blocks.
The reason of we separate the environment from the EOSIO's architecture is attacks against components of environment can be easily addressed out of the EOSIO's scope, which consequently leading to a modular and clear abstraction for EOSIO.

\subsection{Comparison between EOSIO smart contract and Ethereum's}

\begin{table}[]
\centering
\caption{Comparison between EOSIO smart contract and Ethereum's}
\begin{tabular}{ccc}
\toprule
                       & \textbf{Ethereum smart contract}       & \textbf{EOSIO smart contract}                                                                                                                   \\ \midrule
\textbf{Source Language}        & Solidity                      & C++                                                                                                                                    \\
\textbf{Target Language}        & EVM Bytecode / EWasm (future) & WebAssembly                                                                                                                            \\
\textbf{Amount of Instructions} & 154                           & 172                                                                                                                                    \\
\textbf{Type of Instructions}   & -                             & \begin{tabular}[c]{@{}c@{}}Float number related instructions;\\ Typecast related instructions;\\ \texttt{br\_table} instruction;\end{tabular} \\
\textbf{Control Flow Complexity} & Simple                        & Complicated                                                                                                                            \\
\midrule
\textbf{Running Environment}                      & Ethereum VM                   & EOS VM                                                                                                                               \\
\textbf{Data Structure in VM}    & stack, memory, storage        & stack, local, global, memory, table   \\
\bottomrule
\end{tabular}
\label{table:comparison}
\end{table}

Table~\ref{table:comparison} depicts the comparison between EOSIO smart contract and Ethereum's.
As we can see, though they are written in different programming languages (Solidity and C++), they both store the bytecode in the blockchain. Interestingly, Ethereum also plans to support EWasm~\cite{ewasm}, a variant of WebAssembly, in the future upgrade.
Compared with Ethereum smart contract, EOSIO's is more complicated in both quantity and variety of bytecode instructions.
Specifically, Ethereum contains 32 \texttt{PUSH}, 16 \texttt{DUP} and \texttt{SWAP} to handle different length of data. EOSIO, however, pre-defines four types of data, i.e., \texttt{i32}, \texttt{i64}, \texttt{f32} and \texttt{f64}, to abstract all data structure, including string and struct.
Moreover, EOSIO supports float number related instructions and type conversion between above four data types, which are not supported in Ethereum.
And \texttt{br\_table}, which enables multi forks simultaneously like \texttt{swtich-case}, plays an important role in the control flow of EOSIO smart contract.
According to \cite{he2021eosafe}, the authors claimed that: generally, EOSIO smart contracts is more complicated than Ethereum's in terms of control flow.

Except for the distinction located in smart contracts, their running environments, i.e., virtual machines, also exist some differences.
In Ethereum, both operands and operators are pushed into and popped from the stack. The memory and storage, which adopt simple key-value structure, are used to store temporary and permanent data, respectively.
In EOSIO, the stack plays an identical role. Whether the data is stored in the local or global depends on whether the data can be shared between functions.
The memory and table structures are totally different with the Ethereum VM. Specifically, the former one can be randomly accessible, which is implemented by \texttt{store} and \texttt{load} instructions with pointers; the latter one is used to CRUD permanent data, which is achieved by a multi-index database integrated into the EOSIO (see \S\ref{sec:background:environment}).
Consequently, EOSIO is more complicated than Ethereum in both smart contract and running environment level.

%% file: Section-Vulnerability.tex
\section{Vulnerability}
\label{sec:vul}

\begin{figure*}[tbp]
\centerline{\includegraphics[width=\textwidth]{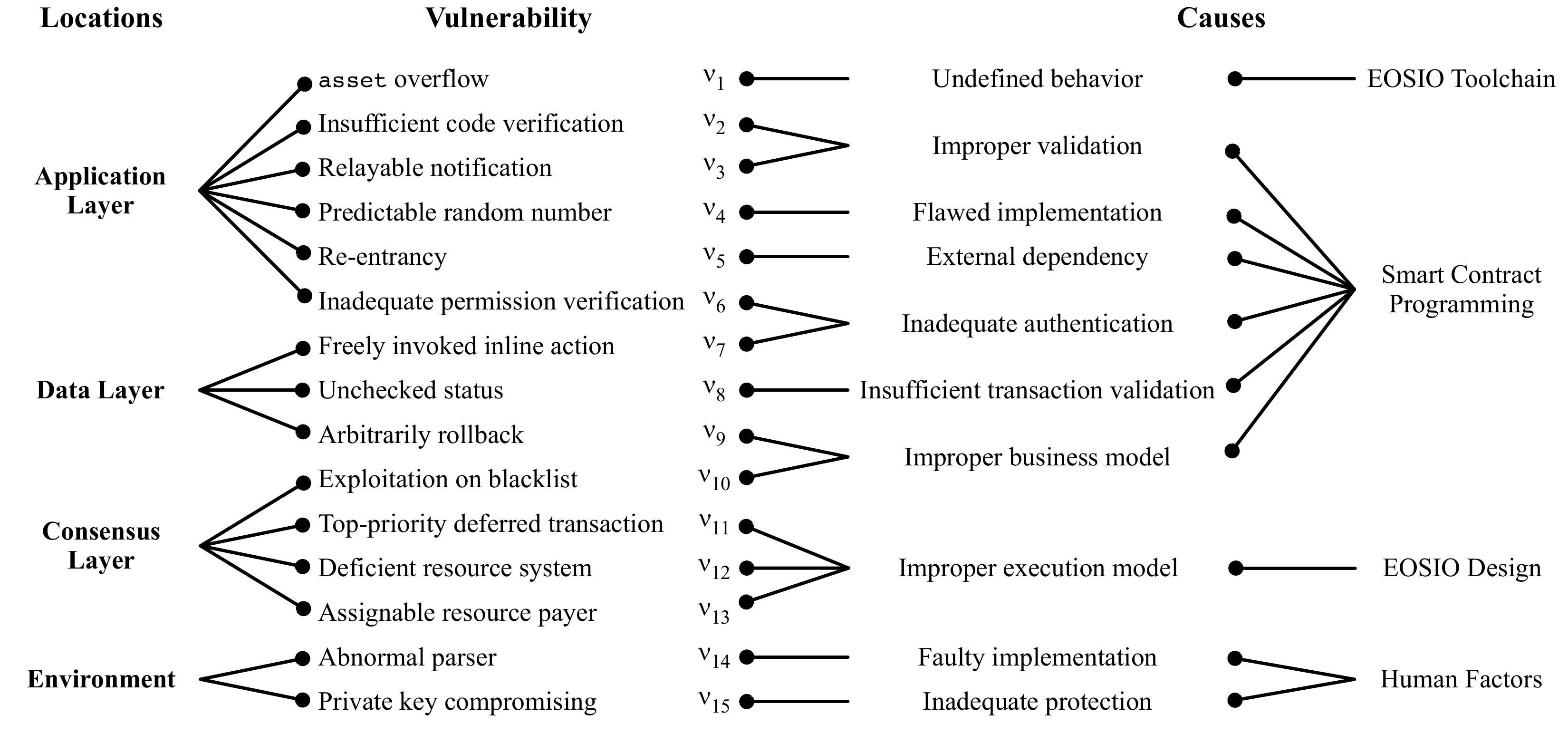}}
\caption{The vulnerabilities in EOSIO, and their corresponding locations and causes.}
\label{fig:vulnerabilities-layer}
\end{figure*}

To cover as many types of vulnerabilities related to EOSIO smart contract as possible, we did a comprehensive investigation from different sources, like  blogs from security companies~\cite{peckshield-blog,slowmist-zone}, forums~\cite{bitcoingambling,eosreddit} and official announcements related to attack events.
After collecting all these information, we manually audit the source code or reverse engineering the WebAssembly bytecode (if the source code is not available) of the corresponding smart contracts.
Consequently, in this section, we discussed all the observed vulnerabilities in the wild, as well as the root cause and the status, i.e., it has already eliminated by official or still open and can be avoided by best practice.
To ease the reference, we adopt notation $\mathcal{V}_i$ to refer to each vulnerabilities (as shown in Fig.~\ref{fig:vulnerabilities-layer}).
In the following, we discuss the vulnerabilities according to its belonged layers.

\subsection{Vulnerabilities in Application Layer}
\label{sec:vul:application}

\subsubsection{\texttt{asset} Overflow ($\mathcal{V}_1$)}
\label{sec:vul:application:asset}

This type of vulnerability is first observed in~\cite{asset-overflow}.
Specifically, EOSIO adopts a user-defined struct, named \texttt{asset}, to encapsulate a certain amount of token. With the uniform definition of \texttt{asset} and its overloaded operators, developers can easily present a piece of money (token) and implement arithmetic operations on the token with the same symbol.
The source code of official implementation of multiplication on \texttt{asset} is shown in Listing~\ref{lst:asset-multiplication}

\begin{lstlisting}[language={C++}, caption={The source code of implementation of multiplication on \texttt{asset}.}, label={lst:asset-multiplication}, mathescape=true]
asset& operator*=(int64_t a) {
  eosio_assert(a == 0 || (amount * a) / a == amount, "multiplication overflow or underflow");
  eosio_assert(-max_amount <= amount, "multiplication underflow");
  eosio_assert(amount <= max_amount, "multiplication overflow");
  amount *= a;
  return *this;
}
\end{lstlisting}

As we can see from L2, it performs a standard implementation of overflow/underflow verification. Then, it returns the result after guaranteeing it ranges from the minimum and maximum limit.
However, any source code, including the library functions, will be compiled to Wasm bytecode format (see \S\ref{sec:bac:app:sc}) to execute on EOS VM.
Listing~\ref{lst:assert-wasm} shows the corresponding Wasm bytecode of L2.

\begin{lstlisting}[language={C++}, caption={The compiled (with \texttt{-O3} optimization) Wasm bytecode corresponding to L2 in Listing~\ref{lst:asset-multiplication}.}, label={lst:assert-wasm}, escapeinside={(*}{*)}]
(call $eosio_assert
  (i32.const 1)    // always true
  (i32.const 224)  // "multiplication overflow or underflow")
)
\end{lstlisting}

We can see that the first argument of assertion is always true, which invalidates the overflow/underflow check.
The reason behind is when both operands are signed integers, the result of overflowed multiplication is undefined~\cite{mul-undefiend}. Coincidently, some compilers with high-level optimization (like \texttt{-O3} in EOSIO) will generate unexpected behaviors encountering undefined result. The overflow of \texttt{asset} could result in the unexpected consumption of any type of tokens.
As this vulnerability is imported by the mis-implementation of official library contract, it is patched timely~\cite{asset-overflow-patch}.

\subsubsection{Insufficient \texttt{code} verification ($\mathcal{V}_2$)}
\label{sec:vul:application:fake-eos}
Fake EOS is the most representative vulnerability in EOSIO, as it is introduced by the mis-implemented \texttt{apply} function (see \S\ref{sec:bac:data:apply}).
Hence, lots of contracts were attacked against this vulnerability~\cite{eosbet-attack-fake-eos, quan2019evulhunter, huang2020eosfuzzer}.
Specifically, its appearance is due to the neglect or improper verification of argument \texttt{code} in \texttt{apply} function, which can be further divided into two circumstances:

\begin{enumerate}
	\item No validation on the \texttt{code} field. Suppose an account (e.g., \texttt{eosio.token1}) also issued a token called EOS by duplicating the official contract in \texttt{eosio.token}. The malicious user could request the \texttt{transfer} in \texttt{eosio.token1} to transfer the fake EOS to a victim, like the process depicted in Fig.~\ref{fig:transfer-eos}. If the victim's \texttt{apply} happens not to check if the value in \texttt{code} is \texttt{eosio.token}, the official one, the victim would be deceived by a worthless and faked \textit{EOS} token.
	\item Even if developers update the \texttt{apply} as Listing~\ref{lst:apply-example}, which only accepts the direct invocation or a notification from \texttt{eosio.token}, it is still vulnerable. The malicious user could call the \texttt{transfer} function directly, in which way the balance for both of them will not be updated. The invocation will then be forwarded to the \texttt{transfer} function to perform the following logic.
\end{enumerate}

Therefore, the vulnerability can only be prevented by a thorough verification on the \texttt{code}. Specifically, if a contract intends to accept EOS token, both of \texttt{action == transfer} and \texttt{code == eosio.token} have to be guaranteed \textit{simultaneously} or it will be affected by such a vulnerability.

\subsubsection{Relayable notification ($\mathcal{V}_3$)}
\label{sec:vul:application:fake-receipt}
As we mentioned in \S\ref{sec:background:data:notification}, EOSIO innovatively introduces the notification mechanism. However, the unfamiliarity of contract developers with the new mechanism may introduce a vulnerability that is named as \textit{fake receipt} vulnerability, which is observed in~\cite{eosbet-attack-fake-receipt}.
In order to enhance the flexibility of notification mechanism, EOSIO enables the \textit{rebroadcast} of notifications.
But to avoid the way like man-in-the-middle attack~\cite{mitm-attack}, EOSIO forbids the modifications of rebroadcasted notification.
To this end, if two malicious accounts coordinate and transfer EOS token to each other, once the payee received the notification from \texttt{eosio.token}, he can immediately relay the notification to the victim.
As both the \texttt{code} and \texttt{action} are valid, the victim may mistakenly believe that he is the beneficiary of the transfer.
As with the Fake EOS vulnerability, this one can also be prevented by a careful arguments' validation, i.e., examining the \texttt{to} field of the transfer notification to distinguish the actual beneficiary.

\subsubsection{Predictable Random Number ($\mathcal{V}_4$)}
\label{sec:vul:application:random}
Gambling DApps plays an important role in EOSIO's ecosystem~\cite{huang2020understanding}. They heavily depend on pseudo-random number generation (PRNG), which can be used to determine which player wins the final jackpot. 
However, current types of pure on-chained PRNG are all flawed, and are attacked several times~\cite{random-attack-1, random-attack-2, eosdice-random}.
Specifically, EOSIO's PRNG relies on some blockchain state as seeds, e.g., \texttt{current\_time}, \texttt{transaction\_id}, and \texttt{tapos\_block\_num}. The variables like the first two, however, are either accessible to everyone or deterministic, which leads to the predictable result of PRNG.
As for the \texttt{tapos\_block\_num}, it refers to the height of the \textit{reference block} (typically the previous block). Taking advantage of the deferred transaction, a future block's reference block seems unpredictable and is properly used as a seed of PRNG.
It turns out, however, that the variable is predictable regardless of how many times the delay transactions invoked (detailed in \S\ref{sec:attack:app:random}).
To mitigate this vulnerability, on the one hand, developers can follow a more secure PRNG prototype proposed by Daniel Larimer~\cite{idealrandom}; on the other hand, they can utilize (de)centralized oracles~\cite{eosio-oracle-1, eosio-oracle-2, eosio-oracle-3} or even its back-end server to generate random number.

\subsubsection{Re-entrancy ($\mathcal{V}_5$)}
\label{sec:vul:app:reentrancy}
Re-entrancy is a well-studied vulnerability in Ethereum smart contracts~\cite{chinen2020ra, rodler2018sereum}, attacks against this vulnerability are firstly observed in~\cite{eos-reentrancy}.
As with Ethereum, this vulnerability is because the order between checking key variable and performing sensitive operations (like function call) is reversed.
Taking transferring tokens as an example, as shown in Fig.~\ref{fig:transfer-eos}, after both of the participants are notified by the token issuer (step 2 and 3), they can perform following logics within their \texttt{transfer} function (if they have).
Intuitively, the order of above operations is: the payer is notified and the notification is handled by the \texttt{transfer}, then the same processes performed in payee's contract.
However, EOSIO processes all the notifications in advance, which brings in a vacuum period for the payer whose money is spent but the response is not conducted in \texttt{transfer}.
So once the malicious payee receives notifications, he could immediately inlined invoke the function again and repeat the above process.
To mitigate this vulnerability, developers should always guarantee updating status before performing operations.

\subsubsection{Inadequate Permission Verification ($\mathcal{V}_6$)}
\label{sec:vul:application:permission}
As we mentioned in \S\ref{sec:background:data:permission}, every account has its related permission hierarchical structure, and each action instance would carry an authority list.
Therefore, once the function in which it performs sensitive operations, e.g., transferring money and updating permanent data, but not conduct a permission validation, it may be abused by malicious users without authorization.
To be specific, there are three authority validating library functions in EOSIO:
\begin{itemize}
	\item \texttt{require\_auth(usr)}: it validates if the action carries the permission that is equivalent to \texttt{usr}'s \texttt{active} permission;
	\item \texttt{require\_auth2(usr, permission)}: it is more flexible and checks the designated \texttt{permission}, like the \texttt{transfer} permission in Fig.~\ref{fig:permission-authority};
	\item \texttt{has\_auth(usr)}: different from the above two functions that raise exceptions, it returns a boolean value indicating whether the \texttt{usr} has authority or not instead of reverting the transaction immediately if the authentication fails.
\end{itemize}
Therefore, to prevent such a vulnerability, the developer is required to utilize one of these three functions to perform strict authorization validation, especially before any sensitive operations.

\subsection{Vulnerabilities in Data Layer}
\label{sec:vul:data}

\subsubsection{Freely invoked inlined action ($\mathcal{V}_7$)}
\label{sec:vul:data:inline-reflex}
Since EOSIO smart contracts are updatable and developers have no obligation to inform users that the contract is updated.
Therefore, if an updated contract performs malicious behavior, it may result in financial losses for its users.
As we explained in \S\ref{sec:bac:data:tx}, the inlined action inherits the context of its parent action instance, including the carrying permission.
For a malicious contract, it is possible to initiate another inlined action after being invoked by an action with other's \texttt{active} permission. In this way, the newly invoked inlined action inherits the victim's permission to perform operations, e.g., transferring EOS, on behalf of the victim.
The immediate reason behind this vulnerability is that no additional authorizations are required to invoke inlined actions on behalf of others.
Hence, EOSIO officially introduces a preserved permission: \texttt{eosio.code}, which should be granted in advance to indicate that the someone completely believes one's code, even if it can be arbitrarily updated.

\subsubsection{Uncheck status ($\mathcal{V}_8$)}
\label{sec:vul:data:hard-fail}
Similar to the status field in Ethereum's and Bitcoin's transaction, which indicates if the transaction executes successfully, in EOSIO transaction there is also a status field whose possible values include: \textit{executed}, \textit{soft\_fail}, \textit{hard\_fail}, \textit{delayed} and \textit{expired}.
Note that only the `executed' status indicates the validity and irreversibility of a transaction, which can then be considered as a mandatory prerequisite for service providing. If the service provider, like centralized exchanges, does not strictly verify the status field, there may be financial losses for them, like the attack event~\cite{hard-fail-attack}.
The mitigation is obvious, i.e., only accept the transactions with the `executed' status before providing services.

\subsubsection{Arbitrary Rollback ($\mathcal{V}_{9}$)}
\label{sec:vul:data:rollback}
As we mentioned in \S\ref{sec:bac:data:tx}, an action's failure would result in its located transaction's rollback.
Some gambling DApps adopt improper \textit{bet-reveal strategy} which results in malicious reversion from attackers, like the way in~\cite{rollback-blacklist,betdiceadmin-rollback}.
Generally speaking, a round of game playing in gambling DApp consists of: receiving the bet from players, calculating a random number, and returning the jackpot to the winner.
However, if the above three steps are all done within a single transaction, a malicious user could append an implicit action to examine changes on his balance to determine if he is the winner. If he is not, he can deliberately revert the whole transaction by a simple failed assertion. Such a rollback will lead to return of the money previously placed to the attacker's account.
The mitigation for DApp developers is to split bet and reveal logic into separate transactions.

\subsection{Vulnerabilities in Consensus Layer}
\label{sec:vul:consensus}

\subsubsection{Exploitation on Blacklist ($\mathcal{V}_{10}$)}
\label{sec:vul:consensus:blacklist}
EOSIO adopts DPoS as its consensus algorithm, which depends on 21 BPs to validate transactions and construct blocks.
To improve the stability of the network, BP can configure a \textit{blacklist} consisting of users who are considered malicious. Transactions invoked from any of the blacklisted users will be dropped and reverted unconditionally.
Some attackers have maliciously exploited this mechanism, leading to financial losses for some gambling DApps~\cite{rollback-blacklist}.
For example, the blacklisted players can brute-forcely attack a gambling DApp by trying all possible results with initiating transactions, which would be dropped and reverted by the BP.
Thus, as long as they can somehow learn which result is collided with the correct answer, they can initiate another transaction immediately with the answer by another non-blacklisted account.
To mitigate this vulnerability, the gambling DApps should adopt a more secure mechanism, e.g., making the results imperceptible or changing randomly in each round.

\subsubsection{Top-priority deferred transaction ($\mathcal{V}_{11}$)}
\label{sec:vul:consensus:transaction-congestion}
As explained in \S\ref{sec:bac:data:tx}, EOSIO introduces deferred transaction to allow transactions whose executing time longer than 30ms exist for some reasons. To guarantee deferred transactions can be executed on time, EOSIO endows them with highest priority.
Therefore, some malicious accounts deliberately invoked deferred transactions in which they looped endlessly, or even schedule another deferred transaction to make it syncing across BPs.
According to~\cite{tx-congestion-attack}, paralyzing more than 100 blocks would only take 0.4 EOS as mortgaging for resources. During the attack period, most of the blocks can only contain no more than 2 transactions.
As this vulnerability is due to the implementation of EOSIO, the official has patched it by dividing normal transactions and deferred transactions into two pools to guarantee a percentage of normal transactions can at least be executed in each round.

\subsubsection{Deficient Resource System ($\mathcal{V}_{12}$)}
\label{sec:vul:consensus:resource}
In most cases, EOSIO's resource system allows users to freely use or exchange resources by mortgaging EOS with a relative low price. But under extreme resource-shortage situations, like after the online of EIDOS~\cite{eidos}, a spam DApp that exhausts existing resources, current resource system has revealed its flaws.
In EOSIO, there are two mainstream ways to obtain resources: 1) mortgage from the official accounts with a fluctuated price (depending on the total staked amount of EOS); and 2) renting a huge amount of but time-limited (typically 30 days) resources by a relative small amount of EOS (depending on the total amount of not being lent resources) from REX, a lending platform~\cite{rex} which allows accounts to deposit his/her idle EOS and resources to earn interests.
However, the EIDOS project was so popular that has almost drained out all the loanable resources in both official contracts and REX with a reasonable price.
According to statistics~\cite{perez2020revisiting}, the transactions related to EIDOS project has accounted for more than 99\%, and the resource price went exaggeratedly high.
EOSIO official did not make any changes to the resource model.
Fortunately, EIDOS project only lasted for 15 months and automatically ended in January 2021, which eventually led to a return to normal resource prices.

\subsubsection{Assignable resource payer ($\mathcal{V}_{13}$)}
Executing a transaction costs CPU and storing permanent data requires RAM (see \S\ref{sec:background:consensus:resource}).
Both of these resources need to be staked or purchased in advance by transaction initiators.
However, EOSIO officially allows one account to pay for the cost of spending resources on behalf of the other one. In some cases, this can lead to resource hijacking attacks~\cite{cpu-hijack-attack, ram-hijack-attack}, resulting in senseless waste or malicious exploitation for profit.

For CPU resource, EOSIO introduced a new feature~\cite{cpu-update} that allows the first account who authorizes the transaction pays for all the consumed CPU.
Some DApps, thus, attract players to participate by paying CPU for accounts. However, malicious accounts could embed inlined actions into the transaction to perform unexpected behaviors without costs.
As for RAM resource, EOSIO allows accounts to designate the RAM payer when executing the table-related library functions, like \texttt{put}, whose second argument indicates the payer of used RAM. Malicious accounts could drain all RAM belonging to the victim by embedding lots of \texttt{put} functions into a transaction with the victim's \texttt{active} permission.
To mitigate the these two vulnerabilities: 1) disable features introduced in \cite{cpu-update} to pay CPU for others; and 2) use a proxy contract, who has limited RAM resources, to perform actions.

\subsection{Vulnerabilities in Environment}
\label{sec:vul:environment}

\subsubsection{Abnormal parser ($\mathcal{V}_{14}$)}
\label{sec:vul:environment:parser}
This vulnerability is first observed in the attack against DApp eosblue.one~\cite{memo-attack}.
To be specific, there is a memo field in each transaction, which is typically used to carry additional information, such as the dice number when player places a bet.
Some back-end server of gambling DApps utilize this mechanism to parse the player's betting options.
However, if the processing logic of memo parser is vulnerable, it would be attacked by carefully constructed content, e.g., the SQL injection.

\subsubsection{Private key compromising ($\mathcal{V}_{15}$)}
\label{sec:vul:environment:private-key}
This type of vulnerability is firstly observed in the attack~\cite{private-key-attack-1}, and similar cases also reported in~\cite{slowmist-zone}.
After compromising the private key, attacker could perform arbitrary actions, e.g., transferring tokens and initiating transactions, on behalf of the victim without authorization.
This will result in both of the reputational loss and financial loss for the victims or even other accounts and platforms.

\subsection{Discussion}
\label{sec:vul:further}

As we can see from Fig.~\ref{fig:vulnerabilities-layer}, there are 4 root causes for all these 15 vulnerabilities, including: \textit{EOSIO Toolchain}, \textit{Smart Contract Programming}, \textit{EOSIO Design}, and \textit{Human Factors}.

\subsubsection{EOSIO Toolchain}
\label{sec:vul:further:toolchain}
There is only 1 vulnerability related to this cause, that is \texttt{asset} overflow ($\mathcal{V}_1$).
Specifically, as EOSIO smart contract is written in C++, then compiled to Wasm bytecode, and finally executed on EOS VM. EOSIO has to introduce a not mature-enough toolchain to complete the pipeline.
So far, only one bug was introduced by it, i.e., compiling stage when the optimization level is set to \texttt{-O3}.
Fortunately, after the vulnerability is committed to the official them, it has been patched timely~\cite{asset-overflow-patch}.

\smallskip
\noindent
\textbf{Insight 1}: EOSIO's toolchain is well implemented. Only one bug is directly related to it but immediately patched by the official team.

\subsubsection{Smart Contract Programming}
\label{sec:vul:further:sc}
Smart contract programming is the most common root cause for vulnerabilities, relating to 9 vulnerabilities.
Specifically, only re-entrancy ($\mathcal{V}_5$) and uncheck status ($\mathcal{V}_8$) have emerged in Ethereum, all the others are not even appeared in the traditional software programming and introduced by EOSIO firstly.
These vulnerabilities are due to either developers' improper validation on arguments and permissions or a flawed implementation on its business logic.
Therefore, we urge EOSIO smart contracts developers to adopt best practices in programming contracts, especially in handling the innovative mechanisms like notification and inlined/deferred actions carefully.

\smallskip
\noindent
\textbf{Insight 2}: Smart contract programming is the most common root cause for EOSIO vulnerabilities. However, all of them can be eliminated by best practices, thus we urge the developers to be careful in introducing and using these innovative mechanisms.

\subsubsection{EOSIO Design}
\label{sec:vul:further:design}
EOSIO's design results in three vulnerabilities: top-priority deferred transaction ($\mathcal{V}_{11}$), deficient resource system ($\mathcal{V}_{12}$) and assignable resource payer ($\mathcal{V}_{13}$).
Specifically, the first one has been patched by the official in an update by splitting the transaction pool of deferred transactions and normal ones, while the latter two cannot be easily mitigated. The problem would emerge again when encountering a killer DApp.

\smallskip
\noindent
\textbf{Insight 3}: EOSIO imports and adopts some new mechanisms, e.g., DPoS consensus, most of which are recognized as excellent solutions. However, the resource system that was intended to allow low-cost interactions among accounts is proven flawed under extreme situations.
We call for a new robust solution.

\subsubsection{Human Factors}
\label{sec:vul:further:human}
Human factors, e.g., social engineering~\cite{social-engineering}, are also observed in the traditional vulnerabilities.
In EOSIO, because some operations are still cannot be achieved purely on-chain (like random number generation), the security of back-end server, as well as the front-end interface of DApps, still should not be underestimated.
Indeed, some existing attacks were against the vulnerabilities located in the end-server of DApps, like abnormal parser ($\mathcal{V}_{14}$) and private key compromising ($\mathcal{V}_{15}$), thus we also stress on the security issues of off-chained components for DApp developers.

\smallskip
\noindent
\textbf{Insight 4}: Except for the on-chain security issues, the components located in environment should also be attached importance by contracts' developers, or the human factors would also result in severe financial losses.

%% file: Section-Attack.tex
\section{Attack}
\label{sec:attack}

\begin{figure*}[tbp]
\centerline{\includegraphics[width=\textwidth]{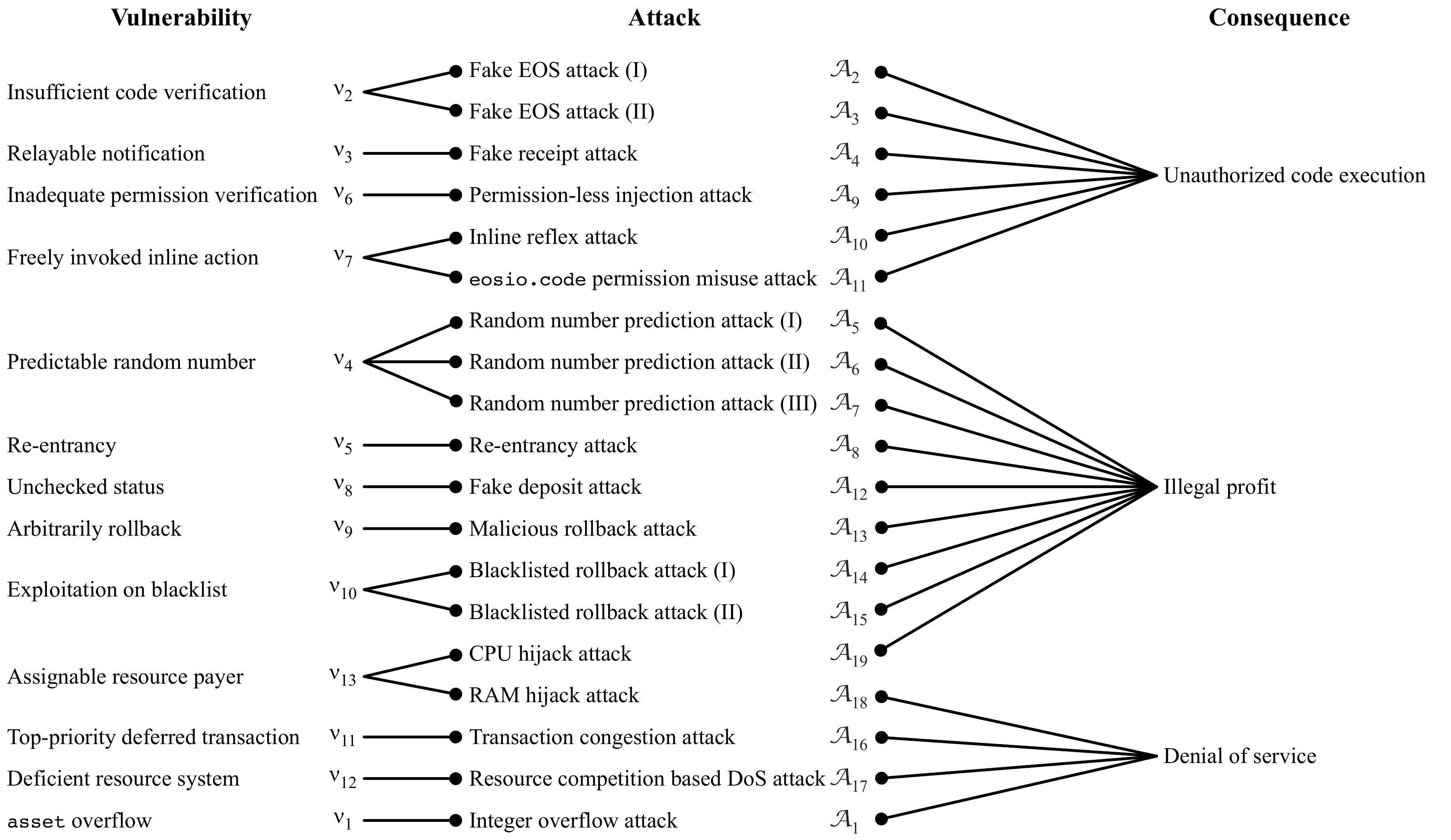}}
\caption{The attacks in EOSIO, and their corresponding vulnerabilities and consequences.}
\label{fig:attacks-layer}
\end{figure*}

In this section, we would present all the attacks related to the vulnerabilities shown in \S\ref{sec:vul}.
Specifically, we will group the attacks, referred as $\mathcal{A}_1$ to $\mathcal{A}_{19}$, according to  positions of the corresponding vulnerabilities.
For each attack, we detail its \textit{victim's background}, \textit{direct impact}, and the \textit{attack tactic}. 
As for the perspicuous relationships between vulnerability, attack and the corresponding impact, we depict them in Fig.~\ref{fig:attacks-layer}.
Note that we omit the attack against vulnerabilities in environment layer, i.e., $\mathcal{V}_{14}$ and $\mathcal{V}_{15}$, because both of them are closely related to human factors that are beyond the scope of this paper.

\subsection{Attacks in Application Layer}
\label{sec:attack:app}

\subsubsection{Integer overflow attack ($\mathcal{A}_1$)}
\label{sec:attack:app:overflow}
\texttt{eosfo.io} is a Fomo3D-like gambling game deployed on EOSIO in account \texttt{zyixjmpxrrpr}.
Specifically, it has a 6-hours countdown, once an arbitrary player purchases a \textit{gem}, the token used in the game, the deadline will be delayed for 30 seconds.
All the EOS tokens that are used to purchase the gems will be accumulated in a prize pool.
If the countdown has successfully decreased as 0, except for the last gem buyer who will get half of the pool, the remaining tokens will be divided proportionally according to the owned gems.
Therefore, once the countdown approaches to 0, a rational player would buy a gem to make him as the final beneficiary. However, such a behavior would also postpone the termination of the game, which will in turn attract other players.

Recall the vulnerability $\mathcal{V}_1$, which may result in integer overflow encountering multiplication of \texttt{asset} struct.
In July 2018, an attacker successfully exchanged more than $9 \times 10^{18}$ gems with $3 \times 10^{-4}$ EOS\footnote{https://www.bloks.io/transaction/09f3fa6668eb4bc3431a7f6df731a9d80856a1f773564489130b7ae043438ab7?tab=traces} by a well-constructed exploitation on $\mathcal{V}_1$.
After the attack transaction, the internal state of the game is messed up, all the following transactions could exchange numerous gems with an extreme low price~\cite{victim-overflow}.
Moreover, the countdown time has underflowed to a meaningless negative number.
Though the official team of \texttt{eosfo.io} announced that the loophole is patched and redeployed the game in account \texttt{eosfoiowolfs}, it was attacked by the same means and suffered more than 60K EOS financial loss~\cite{asset-overflow-result}.

\subsubsection{Fake EOS attack ($\mathcal{A}_2$ and $\mathcal{A}_3$)}
\label{sec:attack:app:fake-eos}
Different from directly transferring tokens between payer and payee, a centralized official account, \texttt{eosio.token}, takes the responsibility of updating the balance table of the whole network. Correspondingly, all accounts can deal with the token transferring request by implementing a simple and fixed \texttt{apply} entry function.
However, improper \texttt{code} validation in \texttt{apply} ($\mathcal{V}_2$) would result in two types of fake EOS attack.
According to~\cite{he2021eosafe, quan2019evulhunter}, these two attacks are prevalent in EOSIO's ecosystem and have already result in extreme financial loss for lots of DApp developers and individual users.

\smallskip
\noindent
\textbf{Attack \#1 ($\mathcal{A}_2$)}
The first type of fake EOS attack is due to the lack of verification on \texttt{code} field in victim's \texttt{apply} function.
For example, the attacker could deploy an account, e.g., \texttt{issuer}, to issue a token that is also named EOS, as EOSIO does not require the uniqueness of token symbol. Then the attacker would call the \texttt{transfer} of \texttt{issuer} who would notify the victim like \textbf{step 3} in Fig.~\ref{fig:transfer-eos}.
If the victim's \texttt{apply} does not examine the notification's initiator (also the issuer of the received token), i.e., the \texttt{code} field, it will be fooled and its \texttt{transfer} function will be invoked unexpectedly.

\begin{lstlisting}[language={C++}, caption={The vulnerable dispatcher of EOSBet.}, label={lst:eosbet-vul-apply}]
void apply(uint64_t receiver, uint64_t code, uint64_t action) {
    auto self = receiver;
    if(action == N(onerror)) {
        eosio_assert(code == N(eosio), "onerror action's are only valid from eosio");
    }
    if(code == self || code == N(eosio.token) || action == N(onerror)) {
        ... // switch to one action
    }
}
\end{lstlisting}

\smallskip
\noindent
\textbf{Attack \#2 ($\mathcal{A}_3$)}
EOSBet~\cite{eosbet} is a famous gambling DApp deployed on EOSIO, whose dispatcher is shown in Listing~\ref{lst:eosbet-vul-apply}.
From ling 6, we can observe that it performed a defensive measure against $\mathcal{A}_2$, i.e., narrowing down the range of acceptable \texttt{code}.
Specifically, the dispatcher only accepts that:
\begin{enumerate}
	\item EOSBet is invoked directly (\texttt{code == self});
	\item the notification comes from \texttt{eosio.token} (\texttt{code == N(eosio.token)});
	\item the notification comes from \texttt{eosio} with the \texttt{onerror} action, indicating an exception should be caught and handled (\texttt{code == N(eosio)} and \texttt{action == N(onerror)})
\end{enumerate}
However, the above three conditions are concatenated by an \textit{OR} operator that is limited by short-circuit evaluation~\cite{short-circuit}.
Thus, an attacker can \textit{directly invoke} the \texttt{transfer} function, which will pass the verification on L6 and into the \texttt{transfer} function, as with the notification is received from \texttt{eosio.token}.
In fact, the attacker \texttt{aabbccddeefg} did perform exactly such a behavior with the manufactured bet in the \texttt{memo} field, thus defrauding the victim\footnote{Bet transaction: https://www.bloks.io/transaction/58ed3541139bba3f91c11c4981052b2d00bfe7ec6cf20208d357c75cc9f943fc}$^,$\footnote{Revealing transaction: https://www.bloks.io/transaction/a630274390fa17e396f6e87b1cdd4a39fff5178992c296d958565b721ccdd3dc}.
Within 15 minutes, the attacker gained more than 44K EOS. 

\begin{figure}[tbp]
\centerline{\includegraphics[width=0.7\columnwidth]{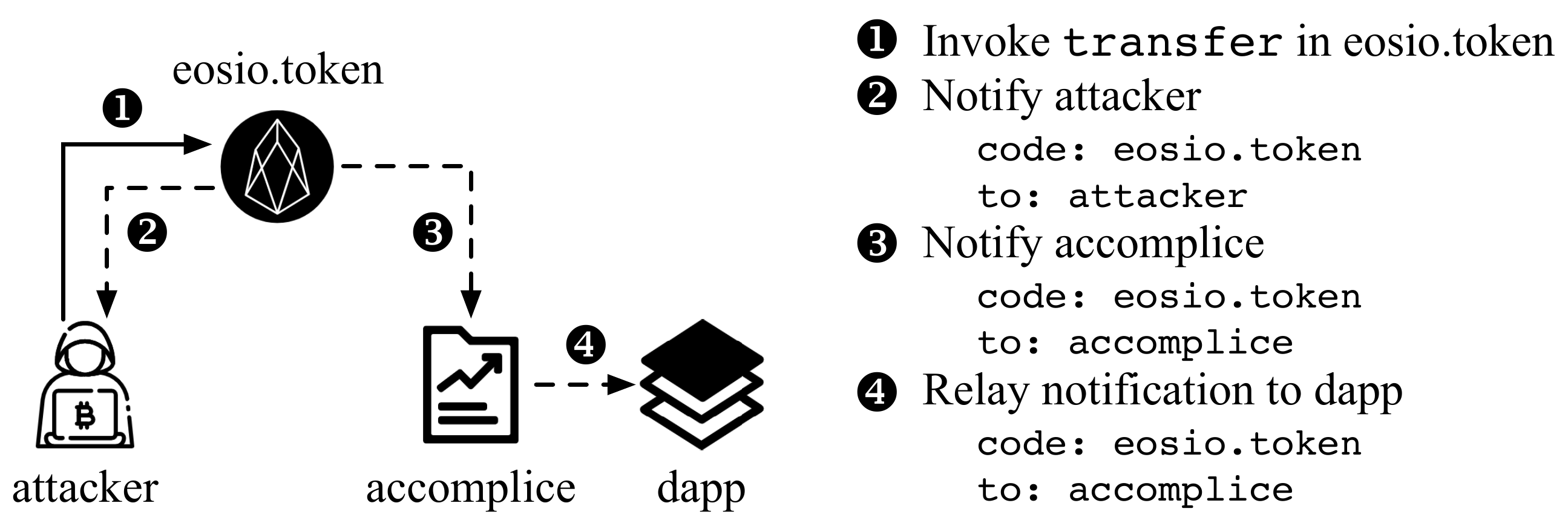}}
\caption{The procedure of fake receipt attack.}
\label{fig:fake-receipt-attack}
\end{figure}

\subsubsection{Fake receipt attack ($\mathcal{A}_4$)}
\label{sec:attack:app:fake-receipt}
After the attack against fake EOS vulnerability, EOSBet, who we mentioned in $\mathcal{A}_3$, was attacked again in October 2018, and suffered more than 142K EOS's financial loss another time~\cite{eosbet-attack-fake-receipt}.
Although the victim's dispatcher was patched timely and immune to the vulnerability $\mathcal{V}_2$, this time the attack exploited $\mathcal{V}_3$ that is in the \texttt{transfer} function, as shown in Listing~\ref{lst:fake-receipt-attack}

\begin{lstlisting}[language={C++}, caption={The problematic \texttt{transfer} function of EOSBet.}, label={lst:fake-receipt-attack}]
void transfer(uint64_t sender, uint64_t receiver) {
  auto transfer_data = unpack_action_data<st_transfer>();
  if (transfer_data.from == _self || transfer_data.from == N(eosbetcasino)) {
  	return;
  }
  // the following operations to deal with the received EOS tokens
}
\end{lstlisting}

From \S\ref{sec:background:data:notification}, we know the transfer notification initiated from \texttt{eosio.token} will be dispatched to EOSBet's \texttt{transfer}.
Within the \texttt{transfer}, EOSBet unpacks the notification and deliberately reverts the notification that is initiated from itself or its official team's contract (L3 to L5).
Then the following operations are performed, like generating a random number once receiving bets.
Note that it only restricts the \texttt{from}, not the \texttt{to}, i.e., the real beneficiary of this transfer.

Taking advantage of $\mathcal{V}_3$, the attacker could perform the attack as shown in Fig.~\ref{fig:fake-receipt-attack}.
Specifically, the \texttt{attacker} transfers EOS to his \texttt{accomplice} through \texttt{eosio.token}. Once the \texttt{accomplice} receives the notification, he will relay the notification immediately to the victim, i.e., \texttt{dapp}, without any change.
As the relayed notification was initiated from the official, it will pass the fake EOS check (including verifying the \texttt{code} and \texttt{action}). However, the \texttt{to} field in \textbf{step 4}, indicating the final beneficiary, is not checked by the EOSBet's \texttt{transfer}.
Thus the fake receipt attack ($\mathcal{A}_4$) can trick the victim into providing services to the attacker by transferring tokens between two accounts who are both controlled by a single entity.

\subsubsection{Random number prediction attack ($\mathcal{A}_5$, $\mathcal{A}_6$ and $\mathcal{A}_7$)}
\label{sec:attack:app:random}
\begin{figure}[tbp]
\centerline{\includegraphics[width=0.7\columnwidth]{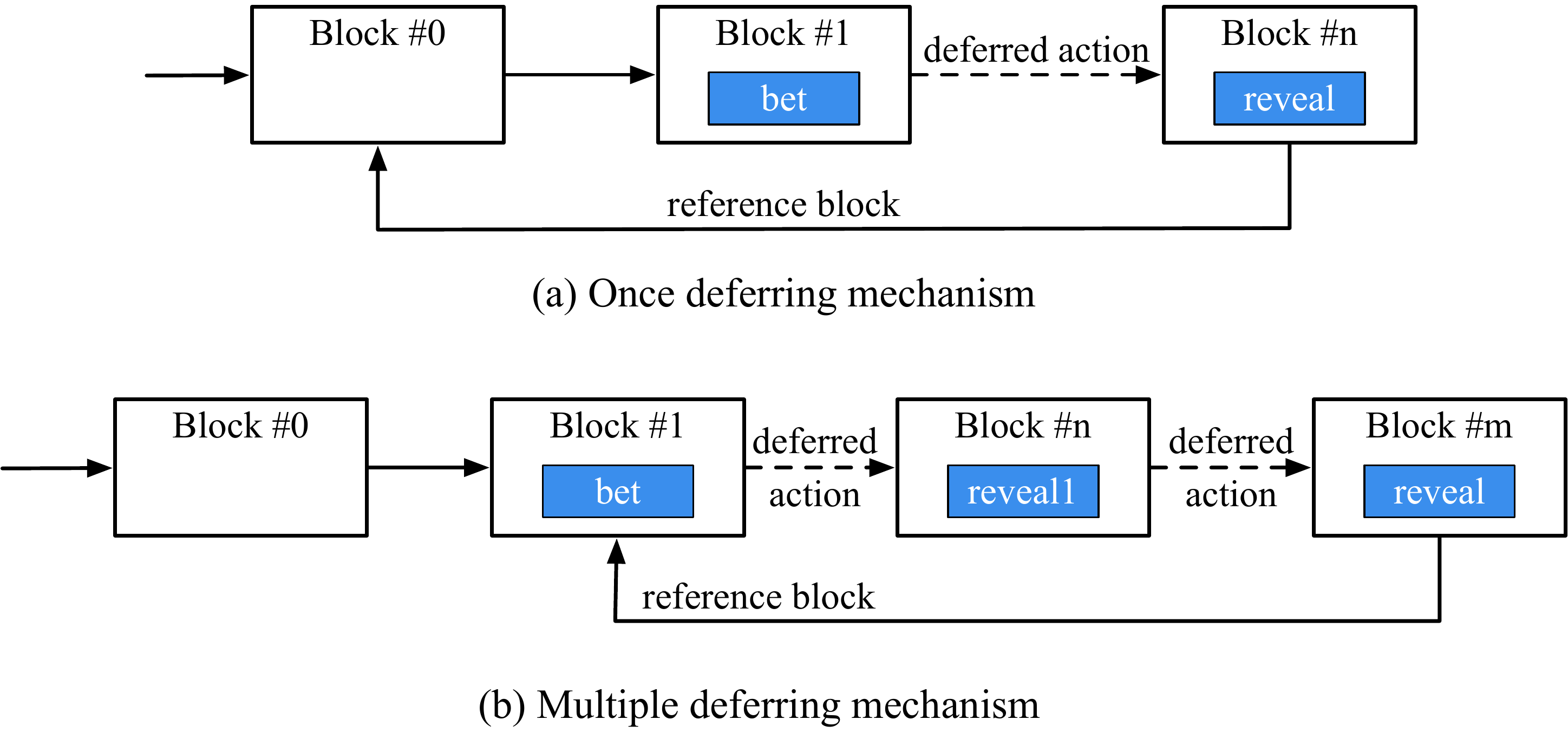}}
\caption{The strategies in revealing random number.}
\label{fig:random-mechanism}
\end{figure}

As we introduced in \S\ref{sec:vul:application:random}, EOSIO's PRNG relies on some blockchain states as seeds and tries to make it \textit{unpredictable} by utilizing some states generated by deferred transactions, like \texttt{tapos\_block\_num}.
For gambling DApps, according to its deferred times between the betting and revealing transactions, we divide the \textit{bet-reveal mechanism} into three categories, as depicted in Fig.~\ref{fig:random-mechanism}

\begin{itemize}
	\item \textbf{Synchronous revealing.} As the name implies, the bet and reveal procedures are located in a single transaction and perform synchronously. However, due to the rollback rule, players can invoke an inlined action (see \S\ref{sec:background:data}) after the revealing process to maliciously revert both the revealing and betting processes to avoid losing money (depicted in $\mathcal{A}_{13}$, see \S\ref{sec:attack:data:rollback}). Therefore, with the adoption of \texttt{tapos\_block\_num}, \textit{once deferring} and \textit{multiple deferring} mechanisms are proposed.
	\item \textbf{Once deferring.} As a special mechanism supported by EOSIO, deferred transaction can be used in revealing jackpot. As depicted in Fig.~\ref{fig:random-mechanism}(a), once the DApp received the bet from players, it will generate a random number and reveal the jackpot in a future block (e.g., \texttt{Block\#n}) by triggering a deferred transaction. Intuitively, \texttt{tapos\_block\_num} in \texttt{Block\#n} is unpredictable as it will point to the latest irreversible block, i.e., the \texttt{Block\#n-1}. However, according to the source code of EOSIO~\cite{tapos-source}, the reference block will be \texttt{Block\#0}, whose metadata can even be retrieved before the \texttt{bet} begins.
	\item \textbf{Multiple deferring.} As the reference block is deterministic when adopting once deferring strategy, the multiple deferring is proposed by some developers. Specifically, as shown in Fig.~\ref{fig:random-mechanism}(b), an additional deferred action, \texttt{reveal1}, is inserted before the \texttt{reveal} function. To this end, the reference block is \texttt{Block\#1}, whose metadata is undeterminated when the betting is performed. 
\end{itemize}

Because the synchronous revealing strategy could be deliberately rollbacked by attackers, most of the gambling DApps adopt once deferring or multiple deferring to complete the bet-reveal process.
Among these DApps, the most well-known one is EOSDice, whose daily volume has reached up to 527K USD in November 2018~\cite{gambling-vol}.
Its huge trading volume and open-sourced seemingly secure strategy attract malicious players. Unfortunately, EOSDice was attacked twice ($\mathcal{A}_5$ and $\mathcal{A}_6$) when it adopted once deferring and multiple deferring strategies, respectively~\cite{eosdice-random}.

 \begin{lstlisting}[language={C++}, caption={The vulnerable random number generating algorithm for EOSDice before the first attack.}, label={lst:eosdice-random1}]
uint8_t random(name name, uint64_t game_id) { 
  asset pool_eos = eosio::token(N(eosio.token)).get_balance(_self, symbol_type(S(4, EOS)).name());
  // seeds
  auto mixd = tapos_block_prefix() * tapos_block_num() + name + game_id - current_time() + pool_eos.amount;
  // generate a number after some bitwise and arithmetic manipulation
  ...
}
\end{lstlisting}

\noindent
\textbf{Attack \#1 ($\mathcal{A}_5$)}
Listing~\ref{lst:eosdice-random1} illustrates the vulnerable code snippet before the first attack, when it adopted once deferring strategy to calculate random number.
Specifically, in the reveal process, the \texttt{random} is invoked to decide the result of dice.
As we can see at L4, the seed \texttt{mixd} is obtained by arithmetic calculation of several parameters. Among these 6 arguments: \texttt{name}, \texttt{game\_id} and \texttt{pool\_eos} (EOS balance of EOSDice) are already determined and can be easily retrieved; \texttt{current\_time} can be calculated by adding the deferred transaction's delay time on the timestamp of the betting transaction; and \texttt{tapos\_block\_prefix} and \texttt{tapos\_block\_num} are the metadata of the block which locates before the betting transaction (see Fig.~\ref{fig:random-mechanism}(a)).
Consequently, the \texttt{mixd} can be calculated and predicted by malicious players, which has resulted in more than 2.5K EOS loss for EOSDice\footnote{https://bloks.io/transaction/a5ce881f87adaa16140183364520c3be251c083fdaededb2295d009aa7b5d94a}.

 \begin{lstlisting}[language={C++}, caption={The vulnerable random number generating algorithm for EOSDice  before the second attack. (This is a simplified version for better display)}, label={lst:eosdice-random2}]
auto eos_token = eosio::token(N(eosio.token));
auto symbol = symbol_type(S(4, EOS)).name();
asset pool_eos         = eos_token.get_balance(_self          , symbol);
asset ram_eos          = eos_token.get_balance(N(eosio.ram)   , symbol);
asset betdiceadmin_eos = eos_token.get_balance(N(betdiceadmin), symbol);
asset newdexpocket_eos = eos_token.get_balance(N(newdexpocket), symbol);
asset chintailease_eos = eos_token.get_balance(N(chintailease), symbol);
asset eosbiggame44_eos = eos_token.get_balance(N(eosbiggame44), symbol);
asset total_eos = asset(0, S(4, EOS));
total_eos = pool_eos + ram_eos + betdiceadmin_eos + newdexpocket_eos + chintailease_eos + eosbiggame44_eos;
\end{lstlisting}

\noindent
\textbf{Attack \#2 ($\mathcal{A}_6$)}
After the first attack, EOSDice inserts an extra deferring like the way shown in Fig.~\ref{fig:random-mechanism}(b).
In this way, the reference block is not determined at the time of betting.
However, EOSDice also updated the process of seed generation, i.e., replacing the \texttt{pool\_eos} by \texttt{total\_eos}, whose calculating process is also detailed in Listing~\ref{lst:eosdice-random2}.
The \texttt{total\_eos} is the sum of EOS balance of some designated accounts, including official accounts, decentralized exchange accounts and other gambling DApps accounts.
Therefore, unlike the simplicity of querying \texttt{pool\_eos}, the \texttt{total\_eos} requires the attacker to predict the balance of these six active accounts, which is impossible.

To achieve the attack, the attacker firstly transfers some EOS tokens to participate the game. Once he is notified by \texttt{eosio.token}, he would immediately imitate the revealing process, i.e., launching a twice deferred transaction with the same delay of EOSDice.
Therefore, at the time of random number calculating, the only parameter can be manipulated is the \texttt{total\_eos} because other metadata of reference block is determined already.
The attacker repeatedly transfers $1 \times 10^{-4}$ EOS (the minimum allowed transferring amount) to any of those accounts till the final result successfully collide with his betting.

\smallskip
\noindent
\textbf{Attack \#3 ($\mathcal{A}_7$)}
Except for the above twice attacks against EOSDice, there still exist some attack instances against flawed random number ($\mathcal{V}_4$).
For example, EOSPlay generates a random number whose seeds are calculated from metadata of a future block and cannot be manipulated directly~\cite{eosplay-attack}.
To carry out the attack, the attacker staked a large amount of CPU resources at a very low price beforehand, and subsequently initiated a huge number of transactions. When these transactions flood into EOSIO network, the decrease of available time in BP results in an abnormally high CPU price. The deviated resource price hampers the desire and increases the difficulty of initiating transactions. The block's body is fully occupied by the attacker's transactions.
Therefore, the metadata of a future block can be predicted to some extent.
This attack had lead to more than 30K EOS loss for EOSPlay~\cite{eosplay-result}.

\subsubsection{Re-entrancy attack ($\mathcal{A}_8$)}
\label{sec:attack:app:reentrancy}
Vaults.sx~\cite{vaults-official} is a decentralized financial DApp on EOSIO, where users can deposit EOS or USDT (a cross-platform token anchoring on USD with enough liquidity) to earn the interest by SXEOS or SXUSDT.
The deposited tokens can further be used in \textit{flashloans}~\cite{flashloan}, i.e., users could loan a large amount of money with a little interest but can only spent within a single transaction.
In May 2021, more than 1.1M EOS and 462K USDT were stolen due to an attack against re-entrancy vulnerability ($\mathcal{V}_5$)~\cite{vaults-attack}.

\begin{lstlisting}[language={C++}, caption={Part of the \texttt{transfer} function in vault.sx.}, label={lst:vault-on-transfer}]
[[eosio::on_notify("*::transfer")]]
void sx::vaults::on_transfer(name from, name to, asset quantity, string memo) {
  const extended_asset out = calculate_retire(id, quantity);

  // update internal deposit & supply
  _vault_by_supply.modify(supply_itr, _self, [&](auto& row) {
    row.deposit -= out;
    row.supply.quantity -= quantity;
    row.last_updated = current_time_point();
  });

  // send underlying assets to sender
  transfer(_self, from, out, _self.to_string());
}
\end{lstlisting}

\begin{lstlisting}[language={C++}, caption={Part of the \texttt{update} function in vault.sx.}, label={lst:vault-update}]
[[eosio::action]] void sx::vaults::update(const symbol_code id) {
  // get balance from account
  const asset balance = eosio::token(N(eosio.token)).get_balance(account, sym.code());

  // update balance
  _vault.modify(vault, _self, [&](auto& row) {
    row.deposit.quantity = balance + staked;
    row.staked.quantity = staked;
    row.last_updated = current_time_point();
  });
}
\end{lstlisting}

Specifically, the attack process is as follows:
\begin{itemize}
	\item[Step 1:]The attacker transfers $2x$ EOS tokens to \texttt{vaults.sx}, who returns $2y$ SXEOS tokens back;
	\item[Step 2:]The attacker transfers half of received SXEOS, i.e., $y$ SXEOS, to \texttt{vault.sx} to redeem $x$ EOS;
	\item[Step 3:]The transfer in \textbf{step 2} would notify the payer (attacker) and the payee (\texttt{vault.sx}) in order.
	\begin{enumerate}
		\item[Step 3.1:]Once the attacker received the notification, his \texttt{transfer} will immediately initiate another two inlined actions: 1) calling \texttt{update} function in \texttt{vault.sx}; and 2) redeeming another half of deposited EOS;
		\item[Step 3.2:]Once the \texttt{vault.sx} received the notification, its \texttt{transfer} (shown in Listing~\ref{lst:vault-on-transfer}) will updates the balance of deposited token (L7) before invoking another inlined action to return the EOS (L13).
	\end{enumerate}
	\item[Step 4:]Following the \textbf{step 3.1}, the \texttt{update} function, listed in Listing~\ref{lst:vault-update}, will retrieve the balance first (L3). Because the EOS is not returned yet (followed by the \textbf{step 3.2}), the \texttt{deposit} at L8 will be unexpectedly inflated. To this end, the actual payout the attacker received by redeem is greater than his pro-rata share, i.e., greater than $x$ EOS.
\end{itemize}

Note that, in practice, the process of triggering the \texttt{update} function in \textbf{step 3.1} is more complex, which requires the attacker to initiate a flashloan.
The above process only depicts the attack with EOS token, it is identical to attack with USDT\footnote{https://www.bloks.io/transaction/3e9f27e101adc4488b5d9256c3682e31ec0da8d4df71d9eba0b7ae546eea3888}.
Consequently, the attacker repeatedly performed the attack till draining out the balance within \texttt{vault.sx}.

\subsubsection{Permission-less injection attack ($\mathcal{A}_9$)}
\label{sec:attack:app:permission}
As illustrated in \S\ref{sec:vul:application:permission} and Fig.~\ref{fig:permission-authority}, EOSIO has a complex permission system. Therefore, inadequate permission verification ($\mathcal{V}_6$) may result in permission-less injection attack, like the DEOS Game attack event~\cite{permission-attack}.
Specifically, the attacker invokes the \texttt{transfer} of \texttt{thedeosgames}, the official contract of DEOS Game. As the contract neither comprehensively verifies the \texttt{code} in \texttt{apply} ($\mathcal{V}_2$), nor checks the carrying permission with the incoming action in \texttt{transfer} ($\mathcal{V}_6$), the attacker could repeatedly construct transfer requests where the payer and payee are real tokens holder and the attacker himself, respectively.
In other words, as the lack of permission verification, the attacker transfers assets out on behalf of others.
According to our statistics, the attacker has received around 4.5K DEOS tokens within just one minute.
In addition, N. He et al.~\cite{he2021eosafe} pointed out that there were 183 permission-less injection attacks against more than 140 contracts through analyzing all the on-chain transactions by heuristic strategy.

\subsection{Attacks in Data Layer}

\subsubsection{Inline reflex attack ($\mathcal{A}_{10}$)}
\label{sec:attack:data:inline-reflex}
In December 2018, gambling DApp FastWin was attacked and lost nearly 2K EOS~\cite{inline-reflex}.
This attack exploits the vulnerability $\mathcal{V}_2$ and $\mathcal{V}_7$, and Peckshield Inc. claimed it is a critical infrastructure flaw residing in authorization verifying module that was timely patched by EOSIO official~\cite{inline-reflex-patch}.

\begin{figure}[tbp]
\centerline{\includegraphics[width=0.8\columnwidth]{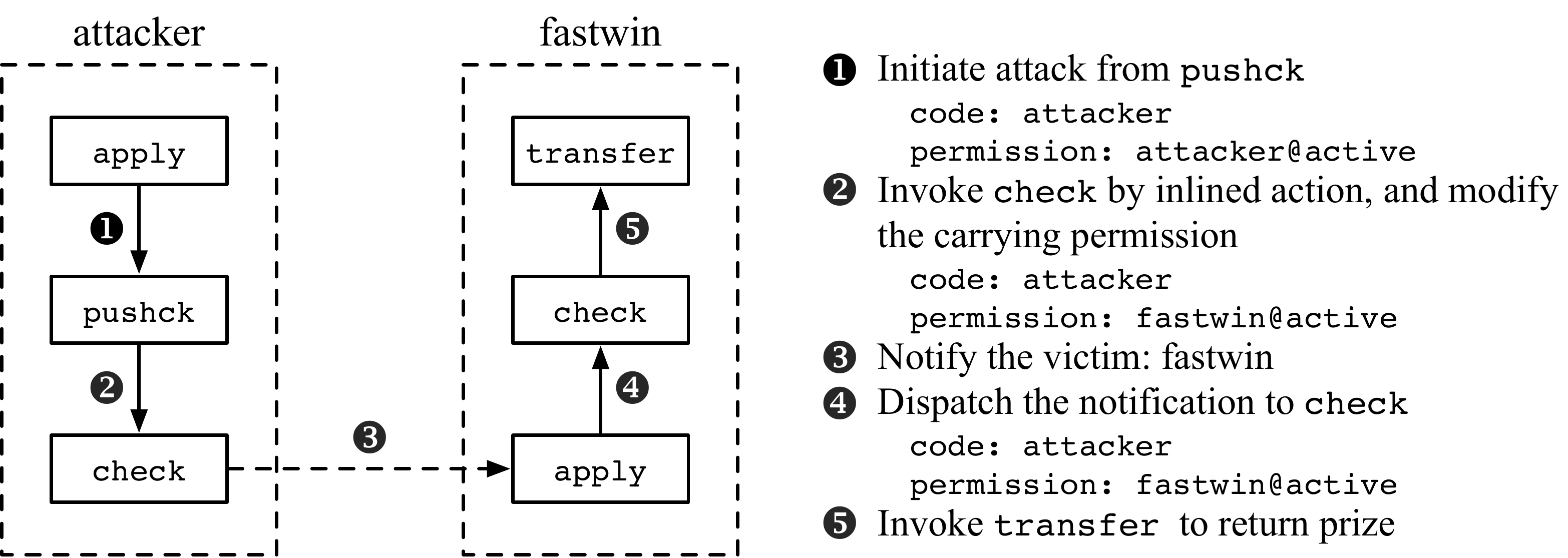}}
\caption{The way of achieving the attack against inline reflex vulnerability.}
\label{fig:inline-reflex}
\end{figure}

According to the attack transaction\footnote{https://eos.eosq.eosnation.io/tx/bc072156b74025ecfdfe9413f7e6d605129bada090aaf2aea4a414b545431e58}, the attack process is illustrated in Fig.~\ref{fig:inline-reflex}.
Specifically, the attacker firstly initiates a transaction by \texttt{pushck} (\textbf{step 1}), in which he invokes his \texttt{check} by an inlined action (\textbf{step 2}).
Two things are worth noting: 1) the victim has a function also named \texttt{check} to decide if a bet wins the jackpot; and 2) the carrying permission is altered to \texttt{fastwin}'s \texttt{active} without authorization in advance.
Then, the attacker actively notifies the victim from the \texttt{check} (\textbf{step 3}). Because the \texttt{code} field happens not to be verified by \texttt{fastwin}'s \texttt{apply}, it dispatches the notification to its own \texttt{check} (\textbf{step 4}).
Before that, EOSIO allows an account impersonate others arbitrarily, so the attacker finally wins jackpots without cost.

\subsubsection{\texttt{eosio.code} permission misuse attack ($\mathcal{A}_{11}$)}
\label{sec:attack:data:eosio-code}
This type of attack also exploits the vulnerability $\mathcal{V}_7$.
Y. Huang et al.~\cite{huang2020understanding} has systematically studied the \texttt{eosio.code} permission misuse behaviors in EOSIO ecosystem, and identified more than 13K related transactions involving around 5.5K accounts.
Generally speaking, if an account actively link someone's \texttt{eosio.code} permission to his \texttt{active} or even \texttt{owner} permission, it means he totally believes in the other's contract, even it can be updated arbitrarily.

Thus, to perform this attack, the attacker needs to lure the victim to link \texttt{attacker@eosio.code} to the \texttt{victim@owner} or \texttt{victim@active}, which is often achieved by adopting social engineering methodologies.
For example, the \texttt{eosfo.io}, the victim in $\mathcal{A}_1$, has deceived its players to link its \texttt{eosio.code} permission to the players' both \texttt{owner} and \texttt{active} permissions for a better availability of the game~\cite{eosio.code-deceive}.
However, if \texttt{eosfo.io} updates the \texttt{transfer} function, e.g., transferring remained EOS to a designated account on behalf of the participated players, the players would suffer a huge financial loss without any warnings.

\subsubsection{Fake deposit attack ($\mathcal{A}_{12}$)}
\label{sec:attack:data:fake-deposit}
Fake deposit attack has been well studied across several blockchain platforms, like Ethereum~\cite{ethereum-fake-deposit} and Ripple~\cite{ripple-fake-deposit}. 
In EOSIO, this attack exploits the vulnerability $\mathcal{V}_8$.
As with other platforms, to perform this attack, the attacker needs to deliberately keep failed transactions on-chain.
Once a service provider does not verify if on-chain transactions failed or not, it will be deceived by such a fake deposit.

In EOSIO, except for the successfully executed transactions, \texttt{hard\_fail} ones are recorded on-chain also.
Those \texttt{hard\_fail} transactions only correspond to the failed deferred transactions~\cite{hard-fail-attack, eosio-fake-deposit}, as they actually take the storage space of BPs before executing.
Additionally, the status indicates \textit{both the transaction and the error handler are objectively failed}, i.e., error handler does not handle the thrown exception properly.
The error handler is often implemented in contract's dispatcher, like L3 to L5 in Listing~\ref{lst:eosbet-vul-apply}.

Take a malicious \texttt{hard\_fail} transaction\footnote{https://eos.eosq.eosnation.io/tx/358286315c70ebee853ace3023909f3ce843c7bbb50803e76be144f566934742} as an example, it is a deferred transaction and initiated by setting delay time through client node manually by the attacker.
But he deliberately transfers unaffordable amount of EOS tokens to the victim.
According to the implementation of the official \texttt{transfer} function (see Listing~\ref{lst:source-transfer-eosiotoken}), the victim would be notified (L6) prior to the balance verification and update (L10 and L11).
Thus, after notifications to both payer and payee, the attacker's transfer will throw an exception due to insufficient balance.
The exception, however, cannot be caught by attacker's client node, which has no smart contract, resulting in a transaction with \texttt{hard\_fail} status.
The victim will be fooled if he only considers the notification from \texttt{eosio.token} but not the transaction's status before providing services.

 \begin{lstlisting}[language={C++}, caption={The implementation of the \texttt{transfer} function in eosio.token.}, label={lst:source-transfer-eosiotoken}]
[[eosio::action]] void token::transfer( /* args */ ) {
  // initialization and authority check
  ...
  // notify payer and payee
  require_recipient(from);
  require_recipient(to);
  // check token's validity
  ...
  // check balance and update balance table
  sub_balance(from, quantity);
  add_balance(to, quantity, payer);
}
\end{lstlisting}

\subsubsection{Malicious rollback attack ($\mathcal{A}_{13}$)}
\label{sec:attack:data:rollback}
This type of attack exploits the vulnerability $\mathcal{V}_9$.
The attacker could intentionally revert a transaction that is unprofitable for him. Thus, it is often observed in gambling DApps, like~\cite{rollback-attack}.
Take a gambling DApp that adopts the synchronous bet-reveal strategy (see \S\ref{sec:attack:app:random}) as an instance.
For harmless players, they will transfer money to the gambling DApp, and pray for a jackpot that is calculated by the revealing function.
For malicious players, however, after the revealing action, they are able to initiate an inlined action to the DApp or the token's issuer (e.g., \texttt{eosio.token}) to query if they won the prize.
According to the results, like the change in balance, they will choose either continue the game or intentionally revert it by a failed assertion.
Eventually, the attacker looks like a super lucky guy who wins prizes in each rounds.

\begin{figure}[tbp]
\centerline{\includegraphics[width=0.7\columnwidth]{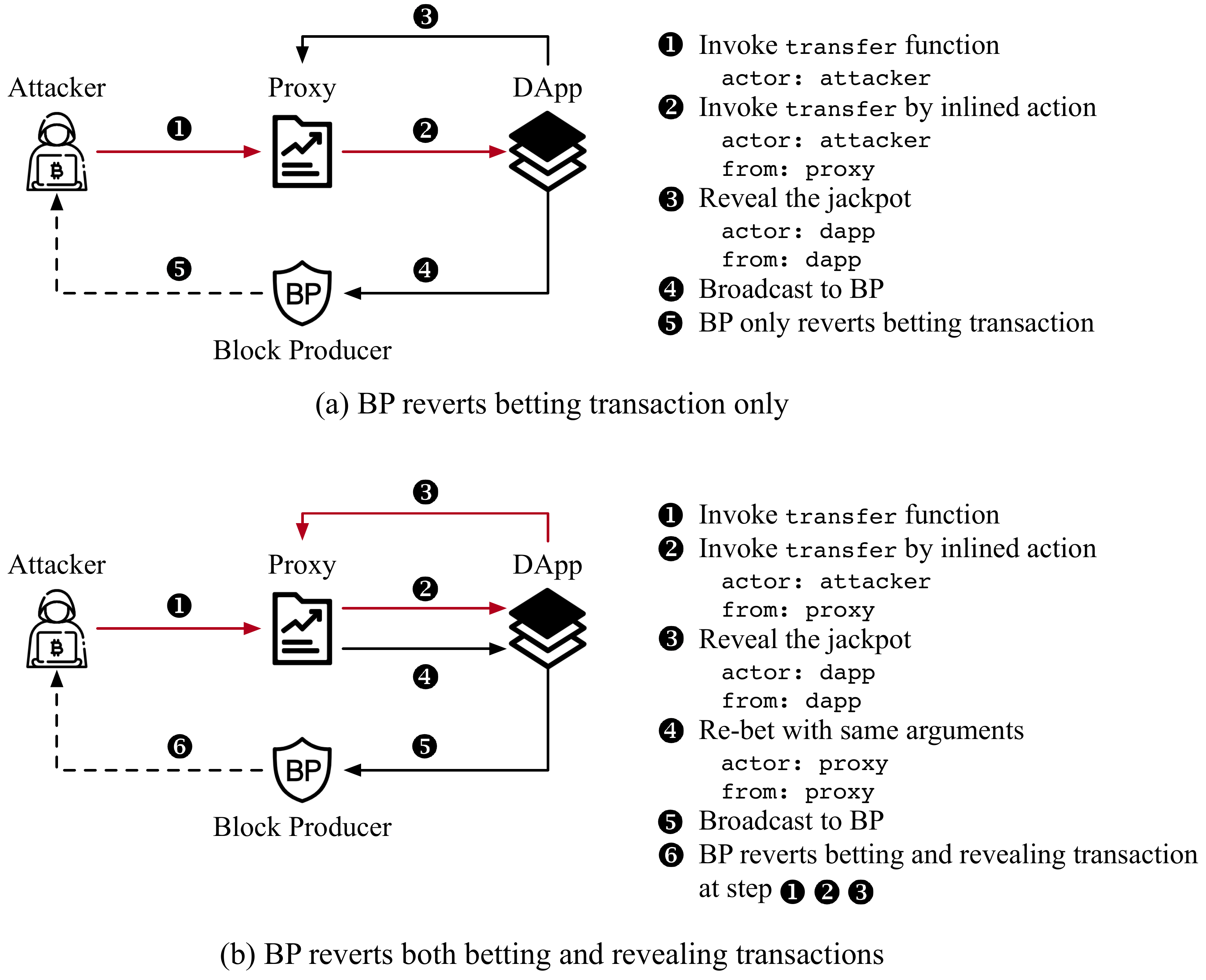}}
\caption{The flowcharts of performing attacks against rollback and replay vulnerabilities in favor of blacklist mechanism in Block Producer.}
\label{fig:blacklist-replay}
\end{figure}

\subsection{Attacks in Consensus Layer}

\subsubsection{Blacklisted rollback attack ($\mathcal{A}_{14}$ and $\mathcal{A}_{15}$)}
\label{sec:attack:consensus:rollback-blacklist}
As we discussed in \S\ref{sec:attack:data:rollback}, the synchronous revealing strategy would lead to attack $\mathcal{A}_{13}$.
Thus, some gambling DApps decide to separate the betting and revealing processes into individual transactions.
Such a seemingly safe strategy not only suffers the random number prediction attack ($\mathcal{A}_5$ to $\mathcal{A}_7$), but also be vulnerable to $\mathcal{V}_{10}$, which takes advantage of the blacklist mechanism.

\smallskip
\noindent
\textbf{Attack \#1 ($\mathcal{A}_{14}$)}
We illustrate the first type of blacklisted rollback attack in Fig.~\ref{fig:blacklist-replay}(a), which requires the isolation of betting and revealing transactions.
Suppose the \texttt{proxy} is created by the malicious \texttt{attacker} who is already blacklisted. The \texttt{proxy} participates in the game through an inlined action (\textbf{step 2}) after receiving the request from the \texttt{attacker} (\textbf{step 1}).
If the bet hits the jackpot, the victim will initiate a new transaction to return the prize to the \texttt{proxy} (\textbf{step 3}), i.e., the actual initiator of the last step's transfer.
According to the consensus algorithm, both of them should be broadcasted to and verified by a BP.
However, the former one will be dropped unconditionally due to the blacklisted initiator, \texttt{attacker}, while only the prize-returning transaction initiated by the \texttt{dapp} is recorded on-chain.
The \texttt{proxy} can keep playing the game with little cost.

\smallskip
\noindent
\textbf{Attack \#2 ($\mathcal{A}_{15}$)}
The $\mathcal{A}_{14}$ takes advantage of the independence between the betting transaction and the revealing one, which in turn urged developers to bind them together. In other words, either transaction's failure will result in the rollback of the both.
However, such a reasonable-sounding strategy is still flawed to $\mathcal{A}_{15}$, whose exploitation is illustrated in Fig.~\ref{fig:blacklist-replay}(b).

Specifically, the first three steps are identical to $\mathcal{A}_{14}$'s.
The loophole hides in \textbf{step 3} that the \texttt{proxy} can determine if he wins or not by the received notification.
If he does, the \texttt{proxy} would immediately initiates a transaction with the same set of arguments (\textbf{step 4}).
Because some gambling DApps use the same set of seed to generate random numbers in such a short period of time, the transaction initiated by a non-blacklisted \texttt{proxy} will 100\% win the jackpot.
Actually, in December 2018, a gambling DApp \texttt{LuckyMe} was attacked by $\mathcal{A}_{15}$'s strategy, and the attacker gained more than 3K EOS with no cost~\cite{slowmist-zone}.

\subsubsection{Transaction congestion attack ($\mathcal{A}_{16}$)}
\label{sec:attack:consensus:transaction-congestion}
This exploitation uses characteristics introduced in $\mathcal{V}_{11}$, i.e., deferred transactions are prioritized over the user-signed normal transactions.
According to Peckshield Inc.'s experiment~\cite{tx-congenstion-eg}, they have nearly paralyzed the whole EOSIO network for more than 2 minutes at a cost of only 0.4 EOS.
Specifically, each deferred transaction needs to be stored till the scheduled time. Once the time is reached, BP will prioritize them to ensure a timely execution and release its storage space.
However, BP will not verify deferred transactions' behavior in advance for efficiency reasons.

To achieve the attack, take the malicious code snippet in Listing~\ref{lst:transaction-congestion-example} as an example, whose only purpose is to perform a dead loop in a deferred transaction (L2).
To this end, the available time slot for a BP would be exhausted if several such malicious transactions are initiated, leading to a denial-of-service situation.
Furthermore, a deferred transaction could invoke other deferred transactions, which may be synced to other BPs.
Consequently, this may result in a prolonged and widespread denial-of-service attack at an extremely low cost.

 \begin{lstlisting}[language={C++}, caption={The way of implementing a malicious deferred transaction by triggering the send function.}, label={lst:transaction-congestion-example}]
[[eosio::action]] void deferred(name from, const string message) {
  while(true) {}
}

[[eosio::action]] void send(name from, const string message, uint64_t delay) {
  eosio::transaction t{};

  t.actions.emplace_back(
    permission_level(from, N(active)),
    _self,
    N(deferred),
    std::make_tuple(from, message));
  
  // set delay in seconds
  t.delay_sec = delay;
  t.send(now(), from);
}
\end{lstlisting}

\subsubsection{Resource competition based DoS attack ($\mathcal{A}_{17}$)}
\label{sec:attack:consensus:eidos}
EOSIO adopts an innovative set of resource model (see \S\ref{sec:background:consensus:resource}), which significantly reduces the cost to initiate transactions.
However, the mortgage price of resources fluctuates significantly under extreme situations (see $\mathcal{V}_{12}$). Normal DApps may suffer DoS attacks due to such a resource competition.
The notorious EIDOS project, who did lead to the above situation, has paralyzed the whole network for up to 15 months and resulted in enormous financial losses~\cite{eidos-result}.

\begin{figure}[tbp]
\centerline{\includegraphics[width=0.8\columnwidth]{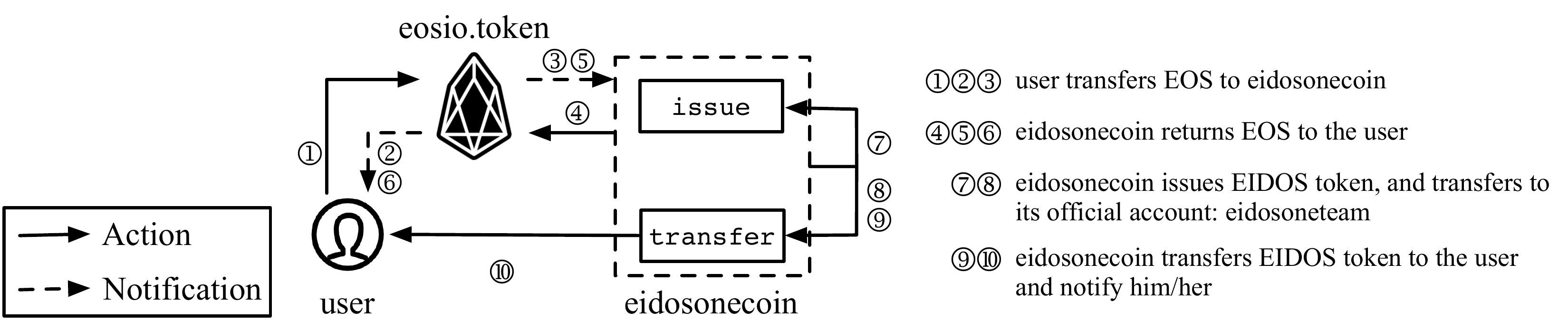}}
\caption{The mechanism of EIDOS project.}
\label{fig:eidos}
\end{figure}

Specifically, EIDOS launched in November 2019 and planned to run for 15 months. Its mechanism is shown in Fig.~\ref{fig:eidos}.
It accepts EOS transfer from investors (\textbf{step 1} to \textbf{3}), and returns it back immediately (\textbf{step 4} to \textbf{6}). Then, it issues some EIDOS token and transfers 0.01\% of accumulated EIDOS tokens to the investor.
In other word, the EIDOS project returns a fixed percent of EIDOS tokens back no matter how many EOS tokens are transferred from an investor.
Hence, the most rational behavior is to transfer $1 \times 10^{-4}$ EOS to EIDOS project as quickly as possible.
The transfer behavior, however, consumes CPU, a limited resource. Combined with the tradable EIDOS token in the secondary market~\cite{eidos-price}, the price of exchanging CPU has skyrocketed.
According to Titan Labs~\cite{cpu-price}, 1 EOS can only exchange 0.08 unit of CPU in December 2019, whose price is 2,000 times than half of the year before.
Such a distorted price of a mandatory resource negatively impacts normal transactions in EOSIO ecosystem.
For example, only one day after its launch, the active users of DApps has dropped 75\% and the transaction volume decreased more than 95\%~\cite{eidos-ps-blog}.

\subsubsection{RAM hijack attack ($\mathcal{A}_{18}$)}
\label{sec:attack:app:ram-hijack}
RAM is a type of resource that is charged when storing permanent data.
For example, \texttt{eosio.token} issues EOS tokens and maintains a balance table for holders. Once a new account claimed some EOS tokens, some RAM will be charged from \texttt{eosio.token}.
According to~\cite{ram-hijack-eg}, in August 2018, an account's RAM was maliciously occupied by a chunk of garbage data, which exploits the vulnerability $\mathcal{V}_{13}$.
An instance of the malicious contract is shown in Listing~\ref{lst:ram-hijack}.

\begin{lstlisting}[language={C++}, caption={An instance of a malicious contract that inserts garbage data.}, label={lst:ram-hijack}]
class dataStorage : public eosio::contract {
    [[eosio::action]] void transfer(name from, name to, asset quantity, std::string memo) {
      // instantiate a multi_index table
      _ttab ttabs(_self,_self);

      for (int i = 0; i < 1000; i++) {
        // The first parameter specifies the account that would pay for any used RAM
        ttabs.emplace(from, [&](auto& data){ 
            // places junk data into the table
            data.id = i;
        });
      }
    }
};

EOSIO_API(dataStorage, (transfer)) //Handles the transfer function
\end{lstlisting}

Specifically, to perform RAM hijacking, the attacker should lure the victim to transfer some EOS tokens to him.
In the attacker's \texttt{transfer}, he will instantiate a table, and repeatedly insert meaningless rows by a native \texttt{emplace} (L8). As the \texttt{emplace} function could designate the payer of the to be used RAM resources, the attacker designates it as \texttt{from}, i.e., the victim who initiated the EOS transfer, his RAM would be occupied suddenly.
According to a experiment~\cite{ram-hijack-attack}, such a 1,000 rows of data would take 100KB of RAM, equivalent to 50 USD.
Occupied RAM cannot be released, nor sold to others, the victim therefore needs to wait for the attacker to actively clear the table and free up RAM resources.

\subsubsection{CPU hijack attack ($\mathcal{A}_{19}$)}
\label{sec:attack:consensus:cpu-hijack}
As we introduced in $\mathcal{V}_{13}$, EOSIO allows the first account who authorizes the transaction to pay for the CPU resources for the following actions~\cite{ONLY-BILL-FIRST-AUTHORIZER}.
Such a behavior would attract players who have no CPU resources to participate in the game, especially during the period with high CPU price.
However, the emergence of EIDOS project, who leads to a significant CPU shortage, has led malicious users into exploiting this feature for profit.

On November 6th, 2019, due to the shortage of CPU resources, the gambling DApp BigGame announced that it would pay the CPU on behalf of players\footnote{https://twitter.com/eosbiggame?lang=en} to attract them.
Thus, some malicious players could pretend to play the game and embed an inlined action in which initiating a transfer to the EIDOS project.
As the CPU resources are payed by the BigGame and EIDOS returns all its received EOS, these players could obtain EIDOS tokens with no cost.

Except for the gambling DApps, the CPU resources in official account \texttt{eosio} can also be hijacked.
Specifically, the function \texttt{bidname} allows accounts to bid a shorter name (less than 12 characters) by transferring EOS tokens. Once someone is overbidden, his bid would be returned by an EOS transfer via \texttt{eosio.token}. As the refund process is initiated by \texttt{eosio}, who owns \textit{unlimited} CPU resources, the malicious account could intentionally be overbidden and embed an inlined action to gain profit from EIDOS project\footnote{https://bloks.io/transaction/0792f37ca3861cede65db056e809f77389057a2a420d5139fa6858e07fd4d505?tab=traces}.

\subsection{Discussion}
As we can see from Fig.~\ref{fig:attacks-layer}, the consequences of all these attacks are divided into three main categories: \textit{Unauthorized Code Execution}, \textit{Illegal Profit}, and \textit{Denial of Service}.
We will discuss them in this section.

\subsubsection{Unauthorized Code Execution}
Unauthorized code execution is a common consequence for attacks in EOSIO ecosystem, e.g., fake EOS attack ($\mathcal{A}_2$ and $\mathcal{A}_3$) and inline reflex attack ($\mathcal{A}_{10}$).
Further, we could divide attacks into two groups: \textit{missing key arguments verification} ($\mathcal{A}_2$, $\mathcal{A}_3$ and $\mathcal{A}_4$) and \textit{missing carrying permission validation} ($\mathcal{A}_{9}$, $\mathcal{A}_{10}$ and $\mathcal{A}_{11}$).
According to \cite{he2021eosafe}, both types of attacks are prevalent on EOSIO and cause significant financial losses, e.g., EOSBet Dice suffered from two attacks, $\mathcal{A}_{3}$ and $\mathcal{A}_4$, which lead to financial losses of 44K and 138K EOS tokens, respectively, around 1M USD~\cite{eosbet-attack-fake-eos, eosbet-attack-fake-receipt}.

\smallskip
\noindent
\textbf{Insight 5}: Missing key arguments verification and carrying permission validation are common on EOSIO platform, which would result in unauthorized code execution and huge amount of financial loss for victim contracts.
We urge developers to strictly limit the possible combinations of input arguments, like requiring \texttt{code == eosio.token} and \texttt{action == transfer} simultaneously.
Moreover, prior to sensitive operations (like invoking other contracts), explicitly validating the carrying permissions is preferable.

\subsubsection{Illegal Profit}
Several attacks could lead to illegal profit consequence, e.g., predicting random number ($\mathcal{A}_5$ to $\mathcal{A}_7$) and blacklisted rollback attacks ($\mathcal{A}_{14}$ and $\mathcal{A}_{15}$).
Among all these related attacks, 6 of them are related to the PRNG process of gambling DApps, the most prevalent category of DApps on EOSIO~\cite{eos-gambling-popular}.
Except for CPU hijack attack ($\mathcal{A}_{19}$), which was first reported in attacks against blockchain platforms, while the other two, re-entrancy attack ($\mathcal{A}_8$) and fake deposit attack ($\mathcal{A}_{12}$), have emerged on other platforms~\cite{the-dao-attack, ethereum-fake-deposit, ripple-fake-deposit}.
However, to achieve successful exploitations, both of them need to take advantage of the introduced by EOSIO, like notification and deferred transaction.

\smallskip
\noindent
\textbf{Insight 6}: Illegal profit would directly import financial loss for DApps' developers. Current attack events have proven the risk of adopting PRNG that purely relies on on-chain seeds. Moreover, some exploitations utilize the new mechanism introduced by EOSIO, like notification and deferred transaction, which should be carefully considered and mastered by developers.

\subsubsection{Denial of Service}
Denial of service, the consequence of some attacks, is fatal for such a distributed system.
All these attacks are related to vulnerabilities that are caused by EOSIO's toolchain ($\mathcal{A}_1$) or EOSIO's design ($\mathcal{A}_{16}$, $\mathcal{A}_{17}$ and $\mathcal{A}_{18}$).
Some of them have been officially patched timely ($\mathcal{A}_1$ and $\mathcal{A}_{16}$) or can be mitigated by best practice ($\mathcal{A}_{18}$).
Unfortunately, there does not exist an effective methodology to mitigate $\mathcal{A}_{17}$, which can be easily reproduced by another spam DApp.

\smallskip
\noindent
\textbf{Insight 7}: EOSIO's toolchain and design have led to several vulnerabilities, against which can be exloited, resulting in denial of service consequence.
EOSIO officially has tried its best to mitigate such a fatal result. Its resource system, especially the fluctuated resources' price calculated by supply and demand relationship, however, is defective under extreme situations.
We counsel EOSIO official to bring forward more efficient and effective solutions.

%% file: Section-Mitigation.tex
\section{Mitigation}
\label{sec:mitigation}
In this section, we would present some mitigations on how to address or avoid vulnerabilities introduced in \S\ref{sec:vul} and denote them as $\mathcal{M}_i$.
Specifically, there exist two mainstream methodologies to automatically identify existing vulnerabilities or attack behaviors: \textit{smart contract analysis} (see \S\ref{sec:mitigation:smart-contract-analysis}) and \textit{transaction analysis} (see \S\ref{sec:mitigation:transaction-analysis}). The former one adopts traditional program analysis methods to identify \textbf{\textit{vulnerabilities}} embedded in EOSIO smart contracts, while the latter one is often used to identify on-chain \textbf{\textit{attack}} transactions.
Both of them can urge developers to examine if their applications are impacted by corresponding vulnerabilities or attacks.
Besides, DApp developers could adopt best practices in programming to actively reduce the possibility of being exploited (see \S\ref{sec:mitigation:programming}).

\subsection{Smart Contract Analysis}
\label{sec:mitigation:smart-contract-analysis}
As an emerging blockchain platform, EOSIO has not been widely noticed by security researchers. However, several work~\cite{he2021eosafe, wang2020wana, huang2020eosfuzzer, quan2019evulhunter, lee2019spent, yan2020tffv} appeared aiming to identify the vulnerabilities in EOSIO smart contracts to facilitate security development for developers.
There are two mainstream smart contract analysis methodologies: \textit{static analysis}~\cite{he2021eosafe, wang2020wana, quan2019evulhunter, lee2019spent, yan2020tffv} and \textit{dynamic analysis}~\cite{huang2020eosfuzzer}.

\subsubsection{Static analysis}
Static analysis is the most prevalent methodology to analyze the EOSIO smart contracts. Specifically, we can divide it into four aspects: \textit{symbolic execution}, \textit{pattern matching}, \textit{formal verification}, and \textit{code audit}.

\paragraph{Symbolic execution}
N. He et al.~\cite{he2021eosafe} ($\mathcal{M}_1$) and D. Wang et al.~\cite{wang2020wana} ($\mathcal{M}_2$) respectively analyzing EOSIO smart contracts by implementing their self-implemented symbolic execution engine.
Specifically, both of the engine extract the control flow graph (CFG) from the Wasm bytecode firstly. Then, engines will traverse all possible paths from the entry function, \texttt{apply}, to obtain semantic information. With the collected constraints along each path, according to the feasibility returned by back-end SMT solvers, researchers can determine if the contract is vulnerable to certain loopholes.
The main advantage of symbolic execution is its theoretically zero false positive under enough computing resources.
To accelerate the analysis, both of them prune infeasible paths in advance with the help of z3, and \cite{he2021eosafe} also adopts some heuristic methodologies to locate suspicious functions.
As for their concerned vulnerabilities, both of them focus on insufficient \texttt{code} verification ($\mathcal{V}_2$) and relayable notification ($\mathcal{V}_3$). The engine proposed by \cite{he2021eosafe} can also identify inadequate permission verification ($\mathcal{V}_6$) and arbitrary rollback ($\mathcal{V}_9$), while the \cite{wang2020wana}'s focuses on a simple version of predictable random number ($\mathcal{V}_4$).

\paragraph{Pattern matching}
L. Quan et al.~\cite{quan2019evulhunter} ($\mathcal{M}_3$) adopts pattern matching to identify insufficient \texttt{code} verification ($\mathcal{V}_2$) and relayable notification ($\mathcal{V}_3$).
Specifically, the tool, EVulHunter, interprets and simulates instructions in each basic block (\texttt{apply}'s and \texttt{transfer}'s) as a simple virtual machine, and simultaneously maintains two vital data structures: \textit{stack} and \textit{memory}. During the simulation, it focuses on some parameters, e.g., \texttt{\_self}, \texttt{from}, and \texttt{to}. Comparing these parameters' logical relationship and pre-defined patterns, EVulHunter is able to identify whether the candidate contract is vulnerable to $\mathcal{V}_2$ and $\mathcal{V}_3$.
However, as the pattern matching heavily relies on the pre-defined patterns, which lack of semantic information and be prone to overfitting problem, the tool may produce false negative cases.

\paragraph{Formal verification}
Z. Yan et al.~\cite{yan2020tffv} ($\mathcal{M}_4$) proposed a tool, TFFV, which is able to transform the source code of EOSIO smart contract into functional equivalent formal verification language.
The above process is implemented by a pipeline consisting of: \textit{lexical analyzer}, \textit{parser}, and \textit{code generator}.
The authors claim that the translation process can be implemented with 100\% accuracy.
Based on the translated formal languages, theoretically, developers are able to prove that the smart contract behaviors exactly as the specification. 
However, the authors did not perform experiments on how to verify if a contract is vulnerable to any of the $\mathcal{V}_i$.

\paragraph{Code audit}
S. Lee et al.~\cite{lee2019spent} ($\mathcal{M}_5$) performed the security analysis by reviewing the source code of EOSIO.
Specifically, they mainly focused on the resource system (CPU and RAM resources) and consensus algorithm (block producer's implementation) of EOSIO.
Through a comprehensive code audit, they have successfully identified assignable resource payer ($\mathcal{V}_{13}$) and top-priority deferred transaction ($\mathcal{V}_{11}$).
Except for the vulnerability identification, they also conducted an experiment to perform related attacks ($\mathcal{A}_{16}$, $\mathcal{A}_{18}$, $\mathcal{A}_{19}$) on testnet to prove EOSIO's inherent loopholes.

\subsubsection{Dynamic analysis}
Except for the mainstream static analysis, dynamic analysis is also adopted by researchers to detect vulnerabilities in EOSIO smart contracts.
Currently, however, there exists only one work \cite{huang2020eosfuzzer} that utilizes \textit{fuzzing testing} to analyze EOSIO smart contracts.

\paragraph{Fuzzing testing}
Fuzzing testing is prevalent in Ethereum smart contract analysis, like \cite{ashraf2020gasfuzzer, jiang2018contractfuzzer}.
In EOSIO, the whole process is similar.
Y. Huang et al.~\cite{huang2020eosfuzzer} presented EOSFuzzer ($\mathcal{M}_6$), a fuzzing framework, to dynamically analyze EOSIO smart contracts.
EOSFuzzer firstly takes both the EOSIO smart contract's bytecode and its corresponding ABI file as inputs, and statically analyzes them to generate a set of input by its pre-defined seeds.
Then, the tool will imitate a malicious agent and initiate transactions according to the randomly generated input.
According to execution logs of targets, it can be easily verified if the target is vulnerable to $\mathcal{V}_2$, $\mathcal{V}_3$ and $\mathcal{V}_4$.

\smallskip
\noindent
\textbf{Insight 8}: All mitigations related to smart contract analysis ($\mathcal{M}_1$ to $\mathcal{M}_6$) and their targeted vulnerabilities are shown in table~\ref{table:mitigations}.
As we can see, none of them focused on the $\mathcal{V}_1$, which can be patched by EOSIO's update.
Most of the tools aim at vulnerabilities resulted from smart contract programming, like $\mathcal{V}_2$ and $\mathcal{V}_3$, which relates to the most prevalent and wide-spread attack events.
However, $\mathcal{V}_5$, $\mathcal{V}_7$, $\mathcal{V}_8$ and $\mathcal{V}_{10}$ are not covered by any of projects.
Among these four vulnerabilities, $\mathcal{V}_7$, $\mathcal{V}_8$ and $\mathcal{V}_{10}$ are partly addressed by EOSIO official or need to be carefully verified by the service providers.
Therefore, we urge security researchers to propose a proper way to handle re-entrancy vulnerability ($\mathcal{V}_5$) and its corresponding attack ($\mathcal{A}_8$).

\begin{table}[t]
\caption{Smart contract analysis mitigations with their corresponding vulnerabilities.}
\centering
\begin{tabular}{ccccccccccccccc}
\toprule
\multicolumn{1}{c|}{} & \multicolumn{1}{c|}{EOSIO Toolchain}   & \multicolumn{9}{c|}{Smart Contract Programming}                                                                                                                                                         & \multicolumn{3}{c|}{EOSIO Design}                                                 &                \\
\multicolumn{1}{c|}{} & \multicolumn{1}{c|}{$\mathcal{V}_{1}$} & $\mathcal{V}_{2}$ & $\mathcal{V}_{3}$ & $\mathcal{V}_{4}$ & $\mathcal{V}_{5}$ & $\mathcal{V}_{6}$ & $\mathcal{V}_{7}$ & $\mathcal{V}_{8}$ & $\mathcal{V}_{9}$ & \multicolumn{1}{c|}{$\mathcal{V}_{10}$} & $\mathcal{V}_{11}$ & $\mathcal{V}_{12}$ & \multicolumn{1}{c|}{$\mathcal{V}_{13}$} & \textbf{total} \\ \midrule
$\mathcal{M}_{1}$ &                   & $\checkmark$      & $\checkmark$      &                   &                   & $\checkmark$      &                   &                   & $\checkmark$      &                    &                    &                    &                    & 4              \\
$\mathcal{M}_{2}$ &                   & $\checkmark$      & $\checkmark$      & $\checkmark$      &                   &                   &                   &                   &                   &                    &                    &                    &                    & 3              \\
$\mathcal{M}_{3}$ &                   & $\checkmark$      & $\checkmark$      &                   &                   &                   &                   &                   &                   &                    &                    &                    &                    & 2              \\
$\mathcal{M}_{4}$ &                   &                   &                   &                   &                   &                   &                   &                   &                   &                    &                    &                    &                    & 0              \\
$\mathcal{M}_{5}$ &                   &                   &                   &                   &                   &                   &                   &                   &                   &                    & $\checkmark$       &                    & $\checkmark$       & 2              \\
$\mathcal{M}_{6}$ &                   & $\checkmark$      & $\checkmark$      & $\checkmark$      &                   &                   &                   &                   &                   &                    &                    &                    &                    & 3              \\ \midrule
\textbf{total}    & 0                 & 4                 & 4                 & 2                 & 0                 & 1                 & 0                 & 0                 & 1                 & 0                  & 1                  & 0                  & 1                  &               \\ \bottomrule
\end{tabular}
\label{table:mitigations}
\end{table}

\subsection{Transaction Analysis}
\label{sec:mitigation:transaction-analysis}
EOSIO, who adopts DPoS as consensus algorithm, would generate transactions thousands of times than Bitcoin and Ethereum within a given period. Therefore, conducting an efficient transaction analysis is challenging and several work~\cite{huang2020understanding, he2021eosafe, zhao2020exploring, zheng2020xblock} has implemented it with different purposes.
For example, with the help of collected transactions, \cite{huang2020understanding, he2021eosafe} detected existing attack events and misbehaviors, while \cite{zhao2020exploring, zheng2020xblock} performed a statistical analysis.

\subsubsection{Transaction-based misbehavior detection}
Executed transactions are recorded on-chain permanently. According to the data stored in transactions (like initiator and carrying permission) and statistical characteristics of transactions (like frequency and trading volume), some work successfully identifies misbehaviors.

To be specific, \cite{huang2020understanding} firstly builds several graphs according to relationships between contracts and accounts in terms of certain types of behavior, like token transferring. Based on these graphs, they calculate a set of statistical metrics via graph analysis.
Then, they propose a clustering algorithm to distinguish \textit{bot accounts} from normal accounts on community level and account level.
They find that these bot accounts are mainly used on bonus hunter and click fraud, and are controlled by DApp teams with different intentions.
Moreover, they propose a heuristic methodology to identify transactions related to attacks $\mathcal{A}_2$ to $\mathcal{A}_5$, $\mathcal{A}_{11}$, $\mathcal{A}_{13}$ and $\mathcal{A}_{16}$, and measure the financial impacts resulted from them.

Except for identifying vulnerabilities through symbolic execution, \cite{he2021eosafe} also proposes a set of heuristic methodology to filter out suspicious attack transactions. They mainly focus on $\mathcal{A}_2$ to $\mathcal{A}_4$, $\mathcal{A}_9$, and $\mathcal{A}_{13}$.
After the filtering process, they further report these attack transactions to the victim DApps to recheck the identifying precision and evaluate the real financial impact.

\subsubsection{Transaction statistics}
\cite{huang2020understanding, zhao2020exploring, zheng2020xblock} perform basic data statistics and analysis on EOSIO's transactions.
For example, both \cite{huang2020understanding} and \cite{zhao2020exploring} build graphs based on the relationships among a certain type of transactions, like account voting and account creation.
Based on graphs, they calculate statistical metrics, e.g., degree distribution, to obtain some macroscopical features from EOSIO network.
Moreover, W. Zheng et al.~\cite{zheng2020xblock} collect transactions from the first 9 million blocks, and divide them by certain behaviors, like transferring and account creation.
For each transaction group, they conduct a basic data statistics work and release them as a public database.

\smallskip
\noindent
\textbf{Insight 9}: The main goal of smart contract analysis is to identify the potential victims, while the analysis based on existing transactions, however, is to filter out the happened but not public-known attack events and misbehaviors, urging self-examination for DApp developers.
Current methodologies heavily rely on heuristic strategies and do not cover all of the attacks we mentioned in \S\ref{sec:attack}. Therefore, a more systematic and comprehensive framework is necessary which is able to put forward more meaningful and interesting insights based on transaction analysis.

\subsection{Smart Contract Programming}
\label{sec:mitigation:programming}

\begin{figure*}[tbp]
\centerline{\includegraphics[width=\textwidth]{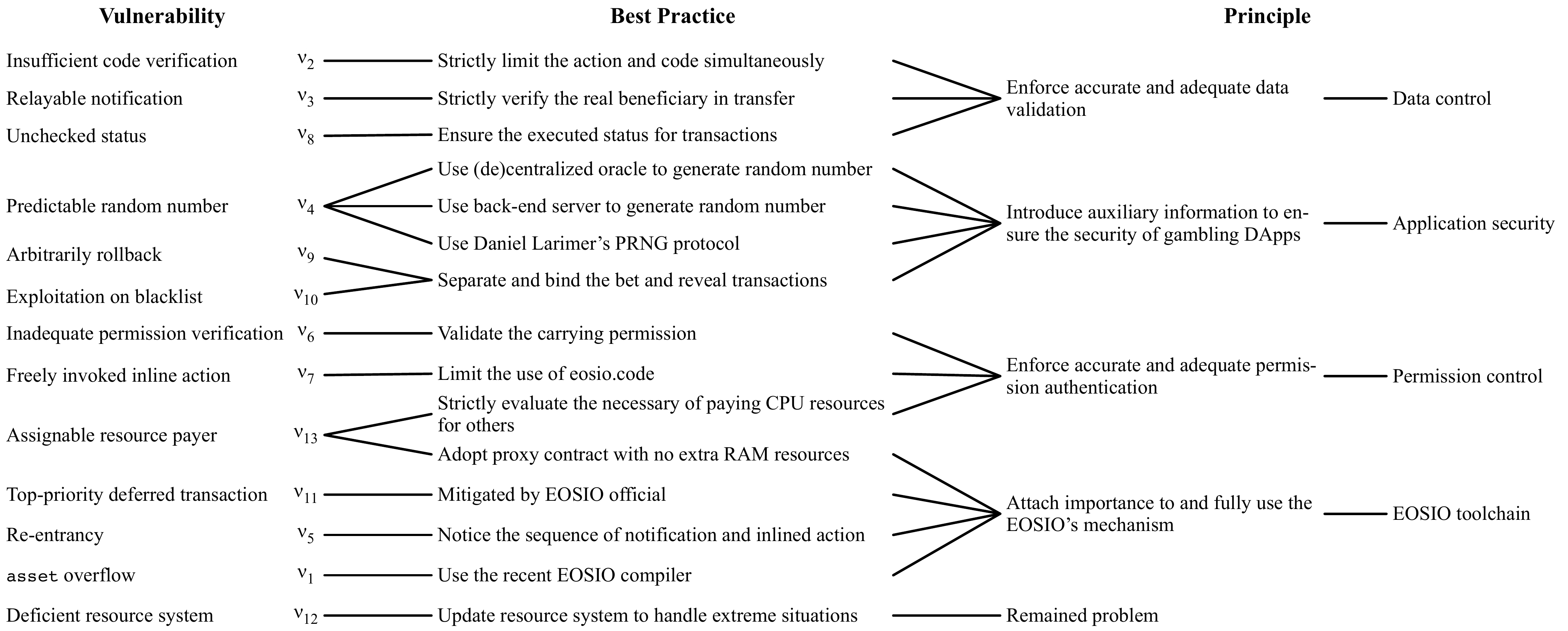}}
\caption{The best practices and behind principles against vulnerabilities in EOSIO.}
\label{fig:mitigations-layer}
\end{figure*}

Except for adopting existing analysis tool to identify vulnerabilities by programming analysis methodologies (see \S\ref{sec:mitigation:smart-contract-analysis}) or extract attack transactions and misbehaviors by transaction analysis (see \S\ref{sec:mitigation:transaction-analysis}), DApp developers can proactively make a defense to these vulnerabilities.
Fig.~\ref{fig:mitigations-layer} illustrates the best practices and their behind principles against vulnerabilities we mentioned in \S\ref{sec:vul} (except for $\mathcal{V}_{14}$ and $\mathcal{V}_{15}$ which is out of EOSIO's scope).
For $\mathcal{V}_2$, $\mathcal{V}_3$ and $\mathcal{V}_8$, whose behind principle is data control, DApp developers could eliminate the possibility of being attacked by strictly verifying values within transactions, like arguments and transaction status.
Similarly, $\mathcal{V}_6$, $\mathcal{V}_7$ and $\mathcal{V}_{13}$ can also be eliminated, whenever accounts and developers carefully hand over or examine carrying permissions.
Additionally, $\mathcal{V}_4$, $\mathcal{V}_9$ and $\mathcal{V}_{10}$ are all related to gambling DApps. These problems can be addressed by introducing extra information, e.g., a price oracle, however, which often means the trade-off between decentralization and reliability for its users.
Remaining ones are all related to EOSIO's mechanism and toolchain. In other words, upgrading to the latest version could partially reduce the harm. However, $\mathcal{V}_{12}$ is still opened and prone to be exploited by another EIDOS-like project.

%% file: Section-Discussion.tex
\section{Discussion}
\label{sec:discussion}

In this section, we discuss the future of EOSIO and give some recommendations, which are all summarized from this work, to developers, official team and security researchers.

\smallskip
\noindent
\textbf{For DApp Developers}
EOSIO is a suitable blockchain platform for DApp developers, as it adopts DPoS consensus algorithm that enables thousand times of TPS than Ethereum, which is currently a mainstream blockchain platform supporting DApp development.
However, EOSIO imports some innovative mechanisms (comparing to Bitcoin and Ethereum), e.g., notification mechanism and resource system.
From \S\ref{sec:vul} and \S\ref{sec:attack}, we could easily conclude that only if DApp developers is knowledgeable about these mechanisms or their projects may suffer huge financial losses.
Therefore, there are some recommendations for DApp developers:
\begin{itemize}
	\item Keep an eye on the release of latest version of EOSIO, especially the official fix of vulnerabilities and the introduction of new mechanism;
	\item Be familiar with the newly imported mechanism, as well as its relationship between the existing ones. For example, the introduction of resource exchange (REX) has affected the resource system to a certain extent~\cite{rex-impact}. Thus, try to postpone the adoption of new mechanism till it is necessary and inevitable;
	\item As for the existing vulnerabilities (see \S\ref{sec:vul}), most of them (especially the ones caused by smart contract programming (see Fig.~\ref{fig:vulnerabilities-layer})) could be mitigated by best practices. For example, $\mathcal{V}_2$ and $\mathcal{V}_3$ can be mitigated by strictly arguments verification;
	\item Mind the emerging attack events, especially the ones whose functionality and purpose are similar/identical to your DApp. With the updatable capability of EOSIO smart contracts, developers should patch their DApps timely or even withdraw them till the immunity of loopholes is ensured by code audit.
\end{itemize}

\smallskip
\noindent
\textbf{For Official Team}
EOSIO official team is trying to communicate with the community positively, like the bug bounty program~\cite{eos-bug-bounty}. However, some vulnerabilities locating in EOSIO system are not discovered by security researchers, but by actual attacks from malicious accounts (like $\mathcal{A}_1$ results from $\mathcal{V}_1$). Thus, it requires the official team to patch it timely to minimize the harm to both EOSIO and DApp developers.
Consequently, some recommendations for EOSIO official team are as follows:
\begin{itemize}
	\item Do not upgrade the system frequently unless it is necessary and urgent. Moreover, strictly audit the code in the new release, it can be performed by specialized third-party blockchain security companies;
	\item According to Occam's razor principle, if it is not necessary, do not import new mechanism. Otherwise, the newly imported one should be decoupled as much as possible with the existing mechanisms, and the instances and best practices of how to use it should be given by the official team;
	\item The official team should consider the update of current resource system, or just eliminate the possibility of emergence of another EIDOS-like project. Its appearance will significantly and negatively impact on platforms's trading volume, users' activity and DApp developers' initiative.
\end{itemize}

\smallskip
\noindent
\textbf{For Security Researchers}
For security researchers, EOSIO smart contracts, i.e., WebAssembly bytecode, is worthy of being studied.
Based on the current work, which discusses the WebAssembly files security~\cite{lehmann2020everything, hilbig2021empirical, stievenart2020compositional}, EOSIO smart contracts analysis~\cite{wang2020wana, he2021eosafe, quan2019evulhunter, huang2020eosfuzzer}, and whether the vulnerable smart contracts are worth being exploited~\cite{perez2021smart}, we propose several recommendations for security researchers:
\begin{itemize}
	\item Apply program analysis methodology (like fuzzing and symbolic execution) on EOSIO smart contracts to explore zero-day vulnerabilities. Then, extract the pattern of the newly discovered loopholes and integrate it into existing analysis tools;
	\item Pay attention to the security of EOSIO platform itself, or its security issue would negatively influence on all deployed smart contracts in terms of security or service availability;
	\item The prosperity of DeFi DApps in Ethereum is likely to be duplicated in EOSIO platform. The attack events against to DeFi DApps, however, heavily depend on victims' flawed business logic, which is hard to be identified and extracted as a generic pattern. Thus, the security analysis towards DeFi DApps need to be covered by future work.
\end{itemize}

%% file: Section-Related.tex
\section{Related Work}
\label{sec:related}

\smallskip
\noindent
\textbf{Survey on Blockchain.}
Actually, there are lots of surveys towards kinds of blockchain platforms~\cite{atzei2017survey,praitheeshan2019security,chen2019survey,li2017survey,wahab2018survey}. Among them, most of which are related to Ethereum. For example, N. Atzei et al.~\cite{atzei2017survey} first finished a survey of vulnerability on Ethereum smart contract. Additionally, they also showed a series of attacks which could exploit these vulnerabilities. 
Except for Ethereum, X. Li et al.~\cite{li2017survey} taken a more macro perspective and did an exhaustive survey of attacks and solutions against the whole blockchain.
As one of the most crucial roles in blockchain, consensus algorithm, has also been fully studied by A. Wahab et al.~\cite{wahab2018survey}.

\smallskip
\noindent
\textbf{EOSIO Security and Ecosystem.}
Because EOSIO is a relatively new blockchain platform, it has not received as much attention as Bitcoin or Ethereum. But there is still some work that analyzes the security and ecosystem of EOSIO~\cite{he2021eosafe,quan2019evulhunter,huang2020understanding,zheng2020xblock,zhao2020exploring,lee2019push,lee2019spent}.
For instance, N. He et al.~\cite{he2021eosafe} has implemented a vulnerability detector which is able to detect four types of loopholes which are included in this survey, and the tool --\textsc{EOSafe} reached a high precision on detection.
Some others also focused on the behavior of account in EOSIO. Y. Huang et al.~\cite{huang2020understanding} achieved a large-scale study to identify the bot accounts with fraudulent activities in EOSIO.
From a higher level, S. Lee et al. has found a threat (that is confirmed by EOSIO official) which could partially freeze the execution of a target smart contract or maliciously consume all the resources of a target user with crafted requests in~\cite{lee2019spent}. Of course, they also discussed possible mitigations against the proposed attacks.